\documentclass[usenatbib,twocolumn]{mnras}
\usepackage{graphicx}
\usepackage{color}
\usepackage{setspace}
\usepackage{amsmath,amssymb,amstext}
\usepackage{dcolumn}
\newcolumntype{d}[1]{D{.}{.}{#1}}
\usepackage[multiple]{footmisc}
\usepackage{multicol}
\usepackage{rotating}

\makeatother
\usepackage{threeparttable}
\usepackage{multicol}
\usepackage{pdflscape}

\def\apj{\rm ApJ}
\def\apjl{\rm ApJL}
\def\apjs{\rm ApJS}
\def\aj{\rm AJ}
\def\jcap{\rm JCAP}
\def\mnras{\rm MNRAS}
\def\nat{\rm Nature}

\def\pasp{\rm PASP}

\def\aap{\rm AAP}
\def\araa{\rm ARA\&A}

\def\lya{Ly$\alpha$}

\def\mgii{\ion{Mg}{\uppercase{II}}}
\def\qsoa{J0313$-$1806}
\def\qsob{J1342$+$0928}

\def\qsod{J0038$-$1527}

\defcitealias{Hennawi2021}{\scshape H2021}

\begin{document}
\title [Mg\,II forest in the EoR]{First measurement of the \ion{Mg}{\uppercase{II}} forest correlation function in the Epoch of Reionization}
\author[Tie et al. ]{Suk Sien Tie$^1$,
     Joseph F. Hennawi$^{1,2}$,
     Feige Wang$^{3}$,
     Silvia Onorato$^{2}$, 
     Jinyi Yang$^{3}$, \newauthor
     Eduardo Ba\~{n}ados$^{4}$, 
     Frederick B. Davies$^{4}$, 
     Jose O\~{n}orbe$^{5}$
    \\
  $^{1}$ Department of Physics, Broida Hall, University of California, Santa Barbara, Santa Barbara, CA 93106-9530, USA \\
   $^{2}$ Leiden Observatory, Leiden University, PO Box 9513, 2300 RA Leiden, the Netherlands \\
   $^{3}$ Steward Observatory, University of Arizona, 933 North Cherry Avenue, Tucson, AZ 85721, USA\\
   $^{4}$Max Planck Institut für Astronomie, Königstuhl 17, D-69117, Heidelberg, Germany \\
   $^{5}$Facultad de Físicas, Universidad de Sevilla, Avda. Reina Mercedes s/n. Campus Reina Mercedes. E-41012, Seville, Spain
   }

\maketitle
\begin{abstract}
In the process of producing the roughly three ionizing photons per atom required to reionize the IGM, the same massive stars explode and eject metals into their surroundings, enriching the Universe to $Z\sim 10^{-3} Z_{\odot}$. While the overly sensitive Ly$\alpha$ transition makes Gunn-Peterson absorption of background quasar light an ineffective probe of reionization at $z > 6$, strong low-ionization transitions like the  \ion{Mg}{II} $\lambda 2796,2804$\AA\, doublet will give rise to a detectable `metal-line forest', if metals pollute the neutral IGM. We measure the auto-correlation of the \ion{Mg}{II} forest transmission using a sample of ten ground based $z \geq 6.80$ quasar spectra probing the redshift range $5.96 < z_{\rm \ion{Mg}{II}} < 7.42$ ($z_{\rm \ion{Mg}{II}, median} = 6.47$).  The correlation function exhibits strong small-scale clustering and a pronounced peak at the doublet velocity  ($\Delta v = 768~{\rm km\,s^{-1}}$) arising 
from strong absorbers in the CGM of galaxies. After these strong absorbers are identified and masked the signal is consistent with noise. Our measurements are compared to a suite of models generated by combining a large hydrodynamical simulation with a semi-numerical reionization topology, assuming a simple uniform enrichment model.  We obtain a 95\% credibility upper limit of $[{\rm Mg/H}] <-3.73$ at $z_{\rm \ion{Mg}{II},median} = 6.47$, assuming uninformative priors on  [Mg/H] and the IGM neutral fraction $x_{\rm{\ion{H}{I}}}$. 
Splitting the data into low-$z$ ($5.96 < z_{\rm \ion{Mg}{II}} < 6.47$; $z_{\rm \ion{Mg}{II},median} = 6.235$) and high-$z$
($6.47 < z_{\rm \ion{Mg}{II}} < 7.42$; $z_{\rm \ion{Mg}{II},median} = 6.72$) subsamples again yields null-detections and  95\% upper limits of $[{\rm Mg/H}] <-3.75$  and $[{\rm Mg/H}] <-3.45$, respectively.  These first measurements set the stage for an approved JWST Cycle 2 program (GO 3526) targeting a similar number of quasars that will be an order of magnitude more sensitive, making the \ion{Mg}{II} forest an emerging powerful tool to deliver precision constraints on the reionization and enrichment history of the Universe.
\end{abstract}


\section{Introduction}
One of the standard probes to study the intergalatic medium (IGM) is neutral hydrogen manifested as a blended collection of absoption lines known as the \lya\ forest. However, due to the large scattering cross section of the \lya\ transition, the absorption lines quickly saturate in the presence of residual neutral hydrogen as low as as 0.01\%, thus rendering it challenging to utilize the \lya\ forest as a probe of the IGM beyond $z \sim 6$. 
An alternative probe of the high-redshift IGM is required. 

Heavy elements, or metals, are expected to be present in the IGM around the Epoch of Reionization due to their ejection by massive stars exploding as supernovae. Low-ionization metals lines (e.g. \ion{O}{I}, \mgii, and \ion{Si}{II}) are good tracers of neutral hydrogen in the pre-reionized IGM due to their similar ionization energy, and their low number densities
mean that they do not saturate as easily as the \lya\ line. Before reionization, forests of these low-ionization lines are expected to arise from predominantly-neutral and overdense regions \citep{Oh2002}. 
During and after reionization, the hardening of the UV background causes the disappearance of the low-ionization forests (e.g. \ion{O}{I}, \ion{Si}{II}, and \ion{Mg}{II}) and gives rise instead to forests of high-ionization lines (e.g. \ion{C}{IV}, \ion{N}{V}, and \ion{O}{VI}). This phase transition signalling reionization has been tentatively observed 
in the circumgalactic medium (CGM) of galaxies (e.g. \citealp{Becker2009,Becker2019,Cooper2019,DOdorico2022}). Absorption lines by metal ions therefore provide us with an additional tool with which to study the IGM and the reionization history. 

Despite the expected presence of metals at early times, their abundance and distribution are unknown, and models do not agree on the typical metallicity of the IGM (e.g. \citealp{OppenheimerDave2006,Oppenheimer2009,Jaacks2018,Kirihara2020,Liu2020}). However, assuming reionization is powered by ionizing photons from massive stars, cosmic reionization and enrichment are then intimately linked, since the same massive stars that reionize the Universe would enrich the IGM to $Z \sim 10^{-3} - 10^{-4} Z_{\odot}$ when they explode as supernovae (e.g., \citealp{EscudeRees1998,Ferrara2016}). Existing observations of metal absorbers concentrate around $z \sim 2-4$, and they suggest a typical IGM metallicity of $Z \sim 10^{-3} - 10^{-2} Z_{\odot}$ as traced by \ion{C}{IV} (e.g. \citealp{SongailaCowie1996,Songaila1997,Schaye2003,Simcoe2011a}) and exhibit a lack of evolution in the metallicity from $z \sim 4.3 - 2$ \citep{Schaye2003}.
Higher-redshift observations of metal absorbers (e.g. \citealp{Simcoe2011b,Ryan-Weber2006,Bosman2017}) are generally probing the CGM of galaxies rather than the diffuse IGM. 
This is because current observations are limited by the use of standard methods that rely on detecting discrete absorbing lines, which necessitates high signal-to-noise ratio (SNR) and high resolution measurements, without which one is likely probing overdense gas around galaxies. Additionally, observing in the near-IR, where metal absorbers redshift into at higher redshifts, faces myriad challenges arising from the higher sky background and worsening detector sensivity and spectral resolution.  

\cite{Hennawi2021} (hereafter \citetalias{Hennawi2021}) introduced
using the auto-correlation of the metal-line forest, which is based on clustering analyses of the \lya\ forest, to detect IGM absorbers. The proposed method treats the metal-line forest as a continuous random field and utilizes all spectral pixels to statistically average down the noise, resulting in far higher sensitivity to the expected weak absorption signals. 
By combining a large hydrodynamical simulation with a semi-numerical reionization topology, \citetalias{Hennawi2021} simulated the \ion{Mg}{II} forest in the pre-reionized IGM at $z=7.5$ and showed that its auto-correlation is sensitive to the neutral hydrogen fraction, $x_{\rm{\ion{H}{I}}}$, and the IGM metallicity, [Mg/H]. The characteristic signal is a peak in the auto-correlation function at the velocity separation of the \ion{Mg}{II} doublet (768 km/s), 
where the peak amplitude increases with [Mg/H] and the large-scale power decreases with increasing
$x_{\rm{\ion{H}{I}}}$. They showed that $x_{\rm{\ion{H}{I}}}$ and [Mg/H] can be constrained to within 5\% and 0.02 dex, respectively, assuming a mock JWST NIRSpec dataset of 10 quasars. \cite{Tie2022} later extended this method to the \ion{C}{IV} forest at $z=4.5$ focusing on the enrichment topology and using a more realistic model for the metal distribution, and showed that the model parameters can also be well-constrained (see Xin et al. in prep for a cross-correlation analysis between metal-line and \lya\ forests). Most recently, \cite{Karacayli2023} introduced an analytical framework for metal absorption based on the halo model and attempted to measure their 1D power spectra in high- and low-resolution data. 

In this work, we aim to jointly constrain the neutral fraction and metallicity of the IGM at $z \geq 6.8$ 
for the first time with the auto-correlation of the \ion{Mg}{II} forest in a sample of ten
high-redshift quasars. We describe our quasar dataset in \S \ref{observations} and the reduction of their spectra in \S \ref{reduction}. In \S \ref{masking}, we address contamination from CGM absorbers that can bias our signal and describe steps to mask them out. We present our results in \S \ref{results}, where we measure the correlation function of the \ion{Mg}{II} forest and describe our procedure
to constrain the metallicity and neutral fraction of the IGM using forward models and Bayesian inference. We conclude in \S \ref{conclusion}. Throughout this work, we adopt a $\Lambda$CDM cosmology with the following parameters: $\Omega_m$ = 0.3192, $\Omega_\Lambda$ = 0.6808, $\Omega_b$ = 0.04964, and $h$ = 0.67038, which agree with the cosmological constrains from the CMB \citep{Planck2020} within one sigma. All distances in this work are comoving, denoted as cMpc or ckpc, unless explicitly indicated otherwise. We define metallicity as $[\mathrm{Mg/H}]$ $\equiv \mathrm{log}_{10}(Z/Z_{\odot})$, where $Z$ is the ratio of the number of magnesium atoms to the number of hydrogen atoms and $Z_{\odot}$ is the corresponding Solar ratio, where we used $\mathrm{log}_{10}(Z_{\odot}) = -4.47$ \citep{Asplund2009}.

\section{Observation}
\label{observations}
Our dataset consists of ten quasars at $z \geq$ 6.80: five $z \geq$ 7 quasars observed with Keck/MOSFIRE \citep{McLean2010,McLean2012}, four $z \geq$ 6.8 quasars observed with Keck/NIRES \citep{nires2004}, and one ultra-deep VLT/XSHOOTER \citep{Vernet2011} observation of a $z=7.1$ quasar \citep{Bosman2017}. Our sample of quasars and their properties are listed in table \ref{tab:qso}. \qsoa\ \citep{Wang2021b} is the current record holder for the highest-redshift quasar discovered, followed by \qsob\ \citep{Banados2018}, while \qsod\ \citep{Yang2019} is a broad-absorption line quasar (as is \qsoa). We also include two new unpublished high-$z$ quasars, denoted as `newqso1' (Ba\~{n}ados et al. in prep) and `newqso2' (Yang et al. in prep) for the rest of this paper.
We note that there are other ground-based observations of $z \geq$ 6.8 quasars, e.g. observed with VLT/XSHOOTER and GEMINI/GNIRS, but these are either too low in signal-to-noise ratio (SNR $\leq$ 3) or have too low spectral resolution in the case of GNIRS. 

Our MOSFIRE observations were executed following an ABAB dither sequence in the $K$-band covering from $1.9 - 2.4$ $\mu$m and through a 0.7" slit, giving a nominal resolution of $R = 3610$ or 
FWHM = 83 km/s, where the spectral sampling is 2.78 pixels per resolution element\footnote{\url{https://www2.keck.hawaii.edu/inst/mosfire/grating.html}}, resulting in a pixel size of $dv = 30$ km/s. The NIRES observations were acquired with an ABBA dither sequence and provide full wavelength coverage of the \textit{YJHK}
bands ($0.95 - 2.45$ $\mu$m) through a fixed 0.55" slit, where the mean resolution is $R = 2700$ or 
FWHM = 111 km/s and the sampling is 2.7 pixels per resolution element\footnote{\url{https://www2.keck.hawaii.edu/inst/nires/spectrometer.html}}, resulting in a pixel size of $dv = 40$ km/s. The XSHOOTER observation of J1120$+$0641 was taken by \cite{Bosman2017} over 30 hours, observed through a 0.9" slit in the VIS and NIR arms. \cite{Bosman2017} measured the resolution to be $R = 7000$ or FWHM = 43 km/s, which we adopt here, and we will use a sampling of 3.7 pixels\footnote{\url{https://www.eso.org/sci/facilities/paranal/instruments/xshooter/inst.html}}, which gives a pixel size of $dv = 12$ km/s.

\begin{table*}
\small
\caption{High-$z$ quasars used in this work} 
\centering 
\begin{minipage}{\textwidth}
    \begin{tabular}{p{0.15cm} p{1.55cm} p{0.55cm} c c c p{0.6cm} c p{0.6cm} c c c c}
        \hline\hline 
         No. & Name & $z_{\rm{em}}$ & $J$ mag & $K$ mag & $t_{\rm{exp}}$ & SNR\footnote{median values calculated from the unmasked regions on the bottom panels of Figures 1 $-$ 6} & SNR$_{\rm{corr}}$\footnote{We discover that the noise vectors produced by the \texttt{PypeIt} pipeline could be slightly over- and under-estimated. We correct for this by applying correction factors to the noise vectors of the quasars as described in the text.} & $\Delta z_{\ion{Mg}{II}}$ & Instr. & $u_k^{\rm eff}$\footnote{Effective weights of quasars towards the correlation function in each redshift bin (see definition following Eqn \ref{eqn:cftwo})} & $u_k^{\rm eff}$ & $u_k^{\rm eff}$\\
         & & & (AB) & (AB) & (sec) & & & & & (All-$z$) & (High-$z$) & (Low-$z$) \\
        \hline
        1 & J0313$-$1806\footnote{\cite{Wang2021b}} & 7.642 & 20.94 $\pm{\,0.13}$ & 19.96 $\pm{\,0.13}$ & 25920 & 18.04 & 18.56 & 1.45 & MOSFIRE & 0.332 & 0.360 & 0.314\\
        2 & J1342$+$0928\footnote{\cite{Banados2018}} & 7.541 & 20.30 $\pm{\,0.02}$ & 20.03 $\pm{\,0.12}$ & 12960 & 14.38 & 15.04 & 1.35 & MOSFIRE & 0.139 & 0.187 & 0.101\\
        3 & J1007$+$2115\footnote{\cite{Yang2020}} & 7.515 & 20.22 $\pm{\,0.18}$ & 19.75 $\pm{\,0.08}$ & 7920 & 9.82 & 10.82 & 1.33 & NIRES & 0.036 & 0.046 & 0.032\\
        4 & J1120$+$0641\footnote{\cite{Mortlock2011}} & 7.085 & 20.35 $\pm{\,0.15}$ & --- & 114000 & 14.7 & 13.88 & 0.91 & XSHOOTER & 0.089 & 0.035 & 0.130\\
        5 & J0252$-$0503\footnote{\cite{Yang2019}} & 7.001 & 20.19 $\pm{\,0.07}$ & 19.92 $\pm{\,0.08}$ & 13680 & 20.63 & 19.0 & 0.82  & MOSFIRE & 0.117 & 0.072 & 0.151\\
        6 & J0038$-$1527\footnote{\cite{Wang2018}} & 7.034 & 19.69 $\pm{\,0.07}$ & 19.33 $\pm{\,0.05}$ & 7200 & 22.86 & 21.67 & 0.83 & MOSFIRE & 0.224 & 0.242 & 0.207\\
        7 & J0411$-$0907\footnote{\cite{Pons2019,Wang2019}} & 6.826 & 20.02 $\pm{\,0.14}$ & 19.56 $\pm{\,0.21}$ & 5760 & 12.58 & 13.53 & 0.66 & NIRES & 0.028 & 0.017 & 0.036\\
        8 & J0319$-$1008\footnote{\cite{Yang2019}} & 6.827 & 20.98 $\pm{\,0.24}$ & --- & 18720 & 7.61 & 8.48 & 0.66 & NIRES & 0.004 & 0.003 & 0.005\\
        9 & newqso1\footnote{Ba\~{n}ados et al. in prep} & 7.0 & --- & 20.39\footnote{Priv. comm.} & 9360 & 5.36 & 6.09 & 0.83 & NIRES & 0.001 & 0.001 & 0.001\\
        10 & newqso2\footnote{Yang et al. in prep} & 7.1 & --- & 19.61\footnote{Priv. comm.} & 5760 & 13.29 & 12.65 & 0.90 & MOSFIRE & 0.028 & 0.036 & 0.024\\
    \hline
    \end{tabular}
\end{minipage}
\label{tab:qso}
\end{table*}

\section{Data Reduction}
\label{reduction}
We reduce our spectra with \texttt{PypeIt}\footnote{\url{https://pypeit.readthedocs.io/en/latest/}}, a Python package for semi-automated reduction of astronomical slit-based spectroscopy \citep{Prochaska2020, Prochaska2020zndo}. After performing standard calibration steps and flux calibration on each exposure, we coadd all exposures of the object and apply telluric correction on the combined spectrum. Since the spectra have different resolutions, wavelength grids, and pixel sizes, we coadd them all onto a common grid with a pixel size of 40 km/s. Finally, we restrict our analyses to a minimum wavelength of 1.95 $\mu$m, corresponding to $z_{\ion{Mg}{II}} = 5.96$. 

\begin{figure*}
\centering
\includegraphics[trim=10 0 0 0, width=0.95\textwidth]{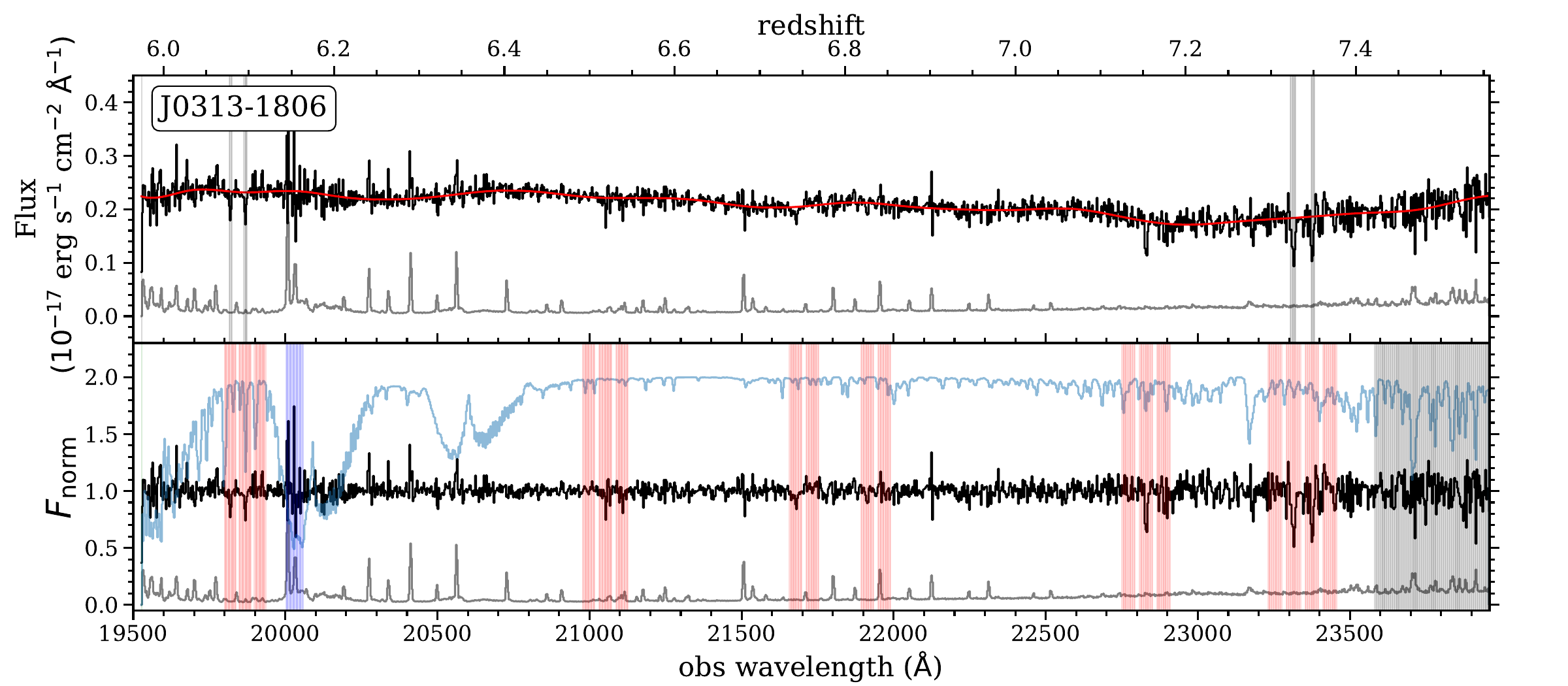}
\includegraphics[width=0.95\textwidth]{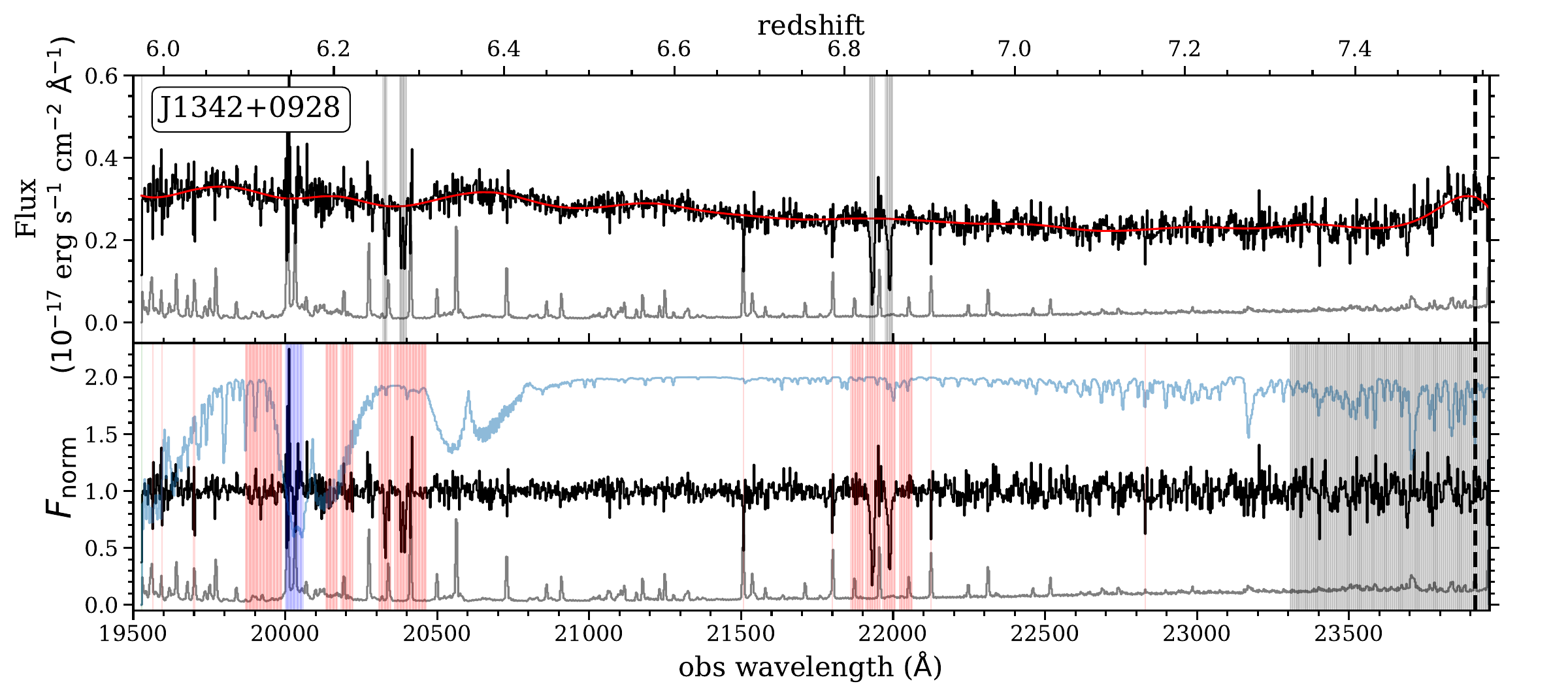}
\includegraphics[trim=0 10 0 0, width=0.95\textwidth]{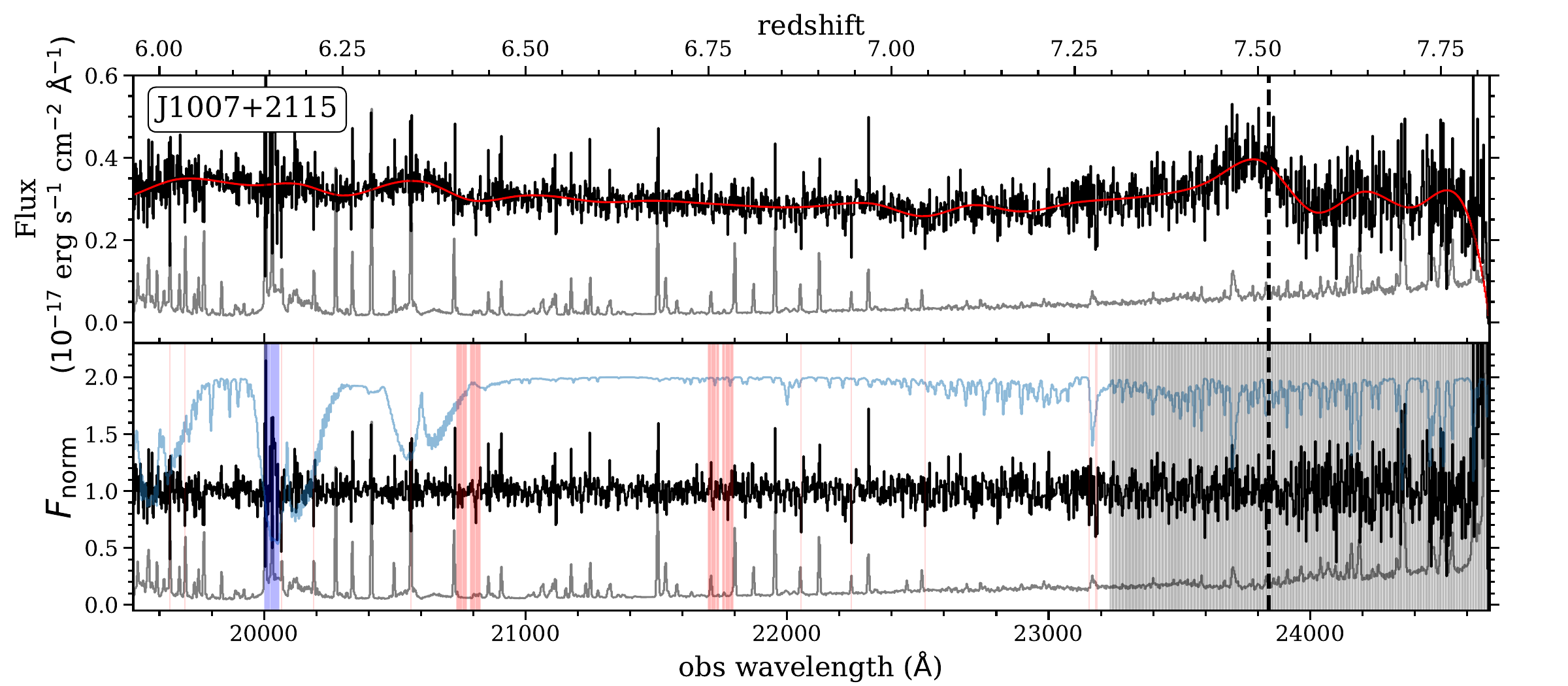}
\caption{Three example spectra from our dataset, J0313$-$1806 (top), J1342$+$0928 (middle), and J1007$+$2115 (bottom). In the top panels of each quasar, the red line  is the fitted quasar continuum, the faint black line is the spectral noise, and the shaded regions indicate masks obtained from \texttt{PypeIt} reduction and from visually-detected strong absorbers, which are applied before continuum fitting. The bottom panel of each quasar shows the continuum-normalized spectrum. Also shown is the telluric spectrum in blue. The multi-colored shaded regions indicate masks due to \texttt{PypeIt} reduction (if available, in green), proximity to quasars and where $z > z_{\rm{QSO}}$ (gray), low SNR due to telluric absorption (blue), and detected CGM absorbers (red; see \S 4), where we use the unshaded regions for the correlation function measurement and inference. The vertical dashed lines indicate the quasar redshift.}
\label{specj0313}
\end{figure*}

\begin{figure*}
\centering
\includegraphics[width=0.95\textwidth]{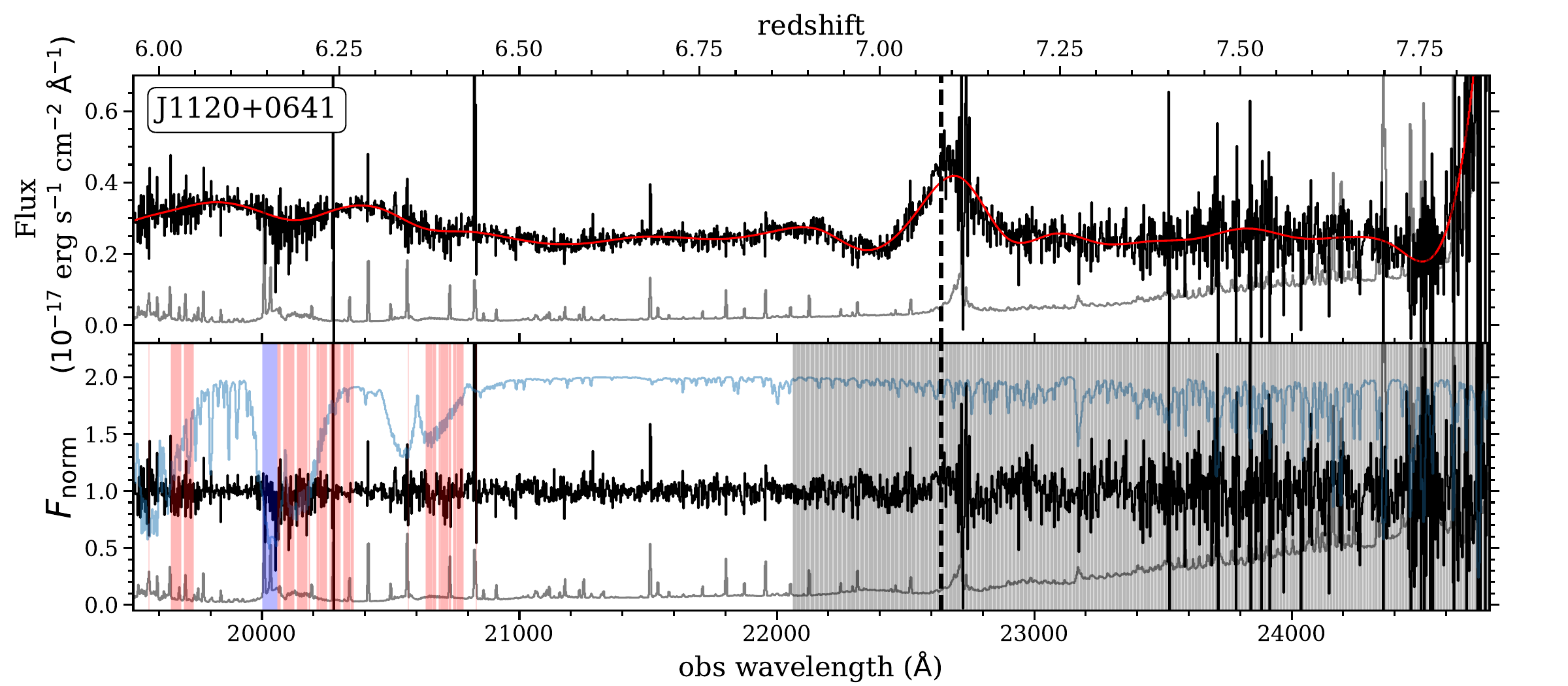}
\includegraphics[width=0.95\textwidth]{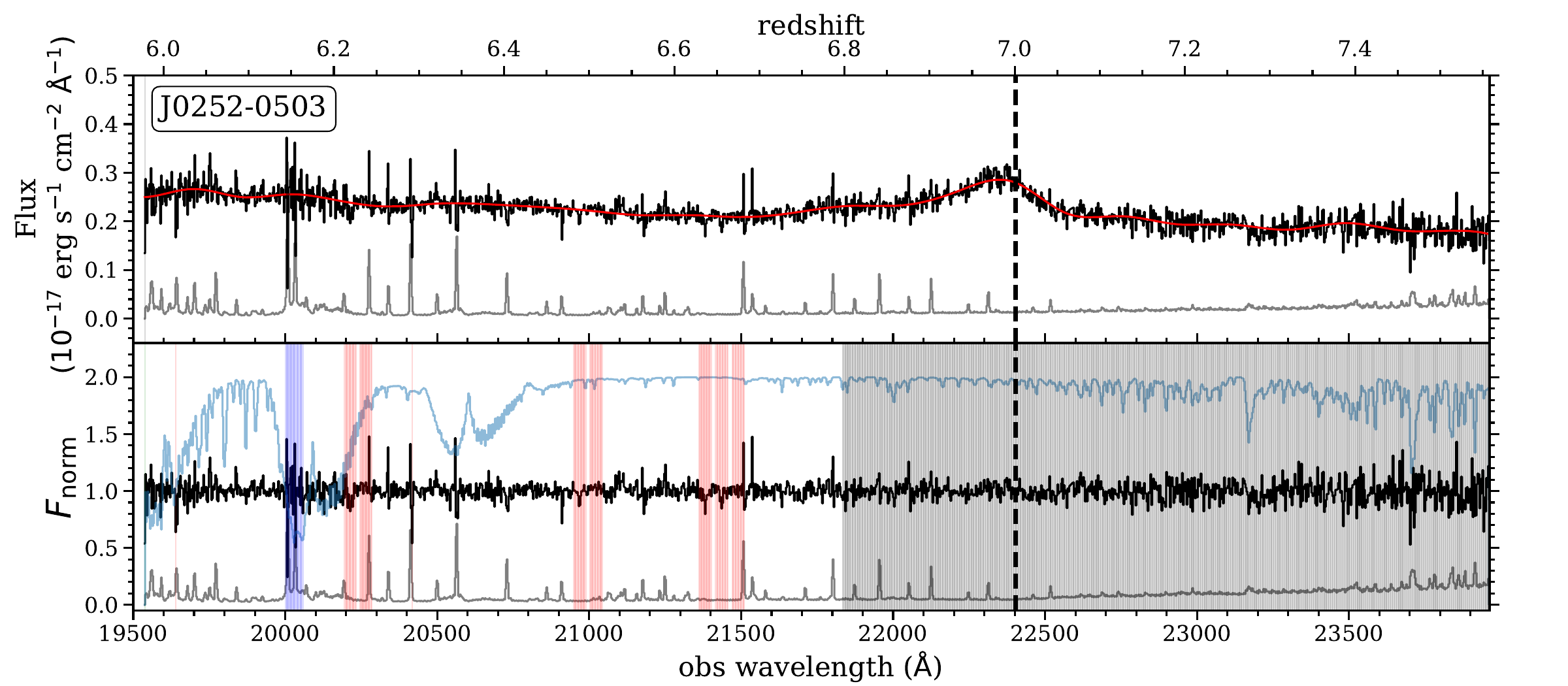}
\includegraphics[width=0.95\textwidth]{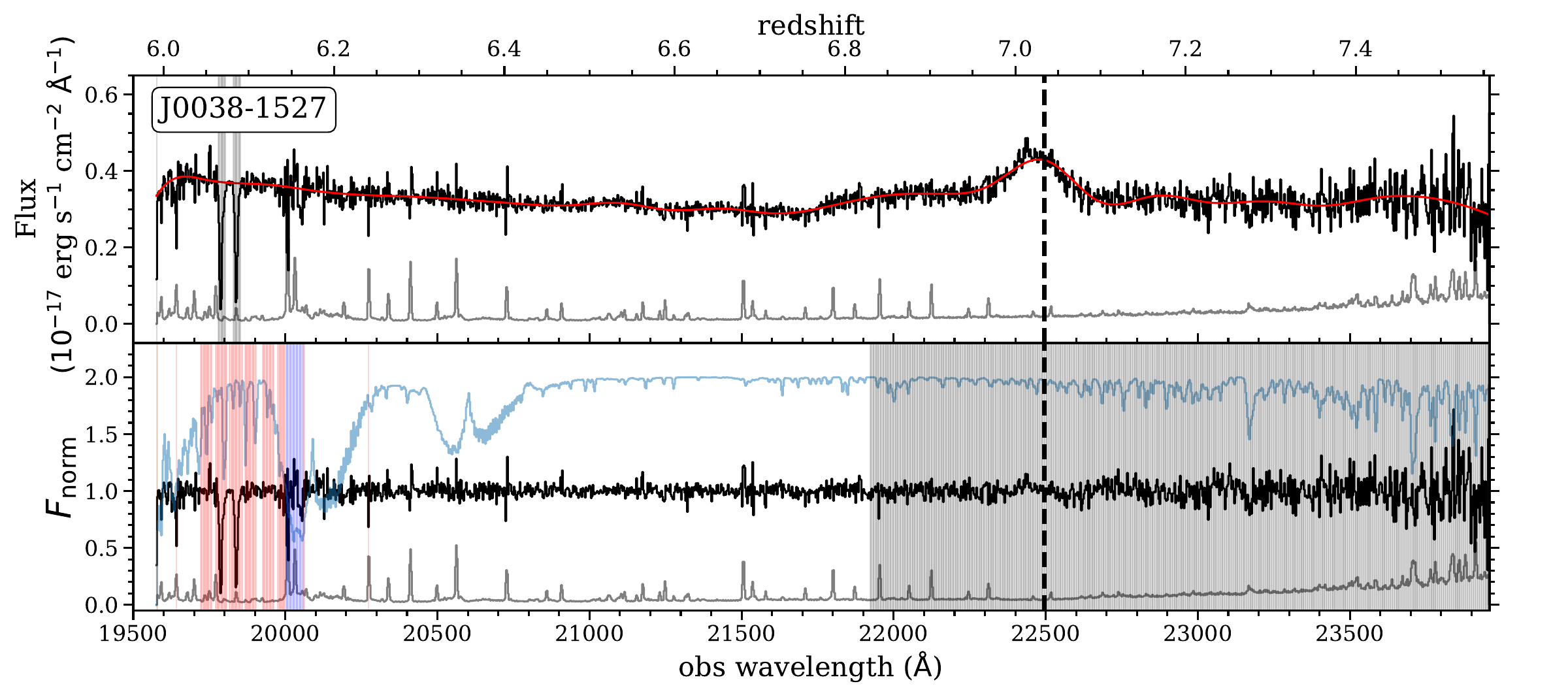}
\caption{Same as Figure \ref{specj0313}, but for J1120$+$0641 (top), J0252$-$0503 (middle), and J0038$-$1527 (bottom).}
\label{specj1120}
\end{figure*}

\begin{figure*}
\centering
\includegraphics[width=0.95\textwidth]{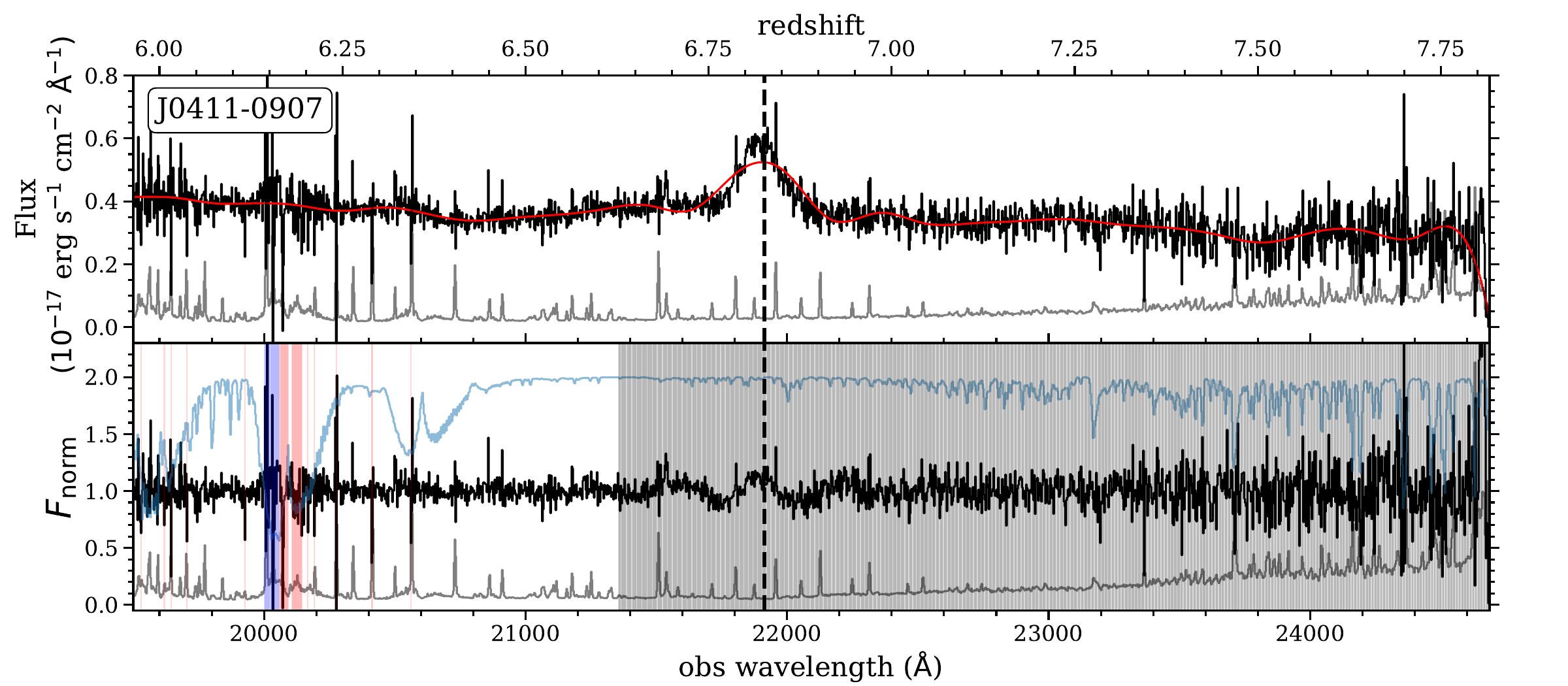}
\includegraphics[width=0.95\textwidth]{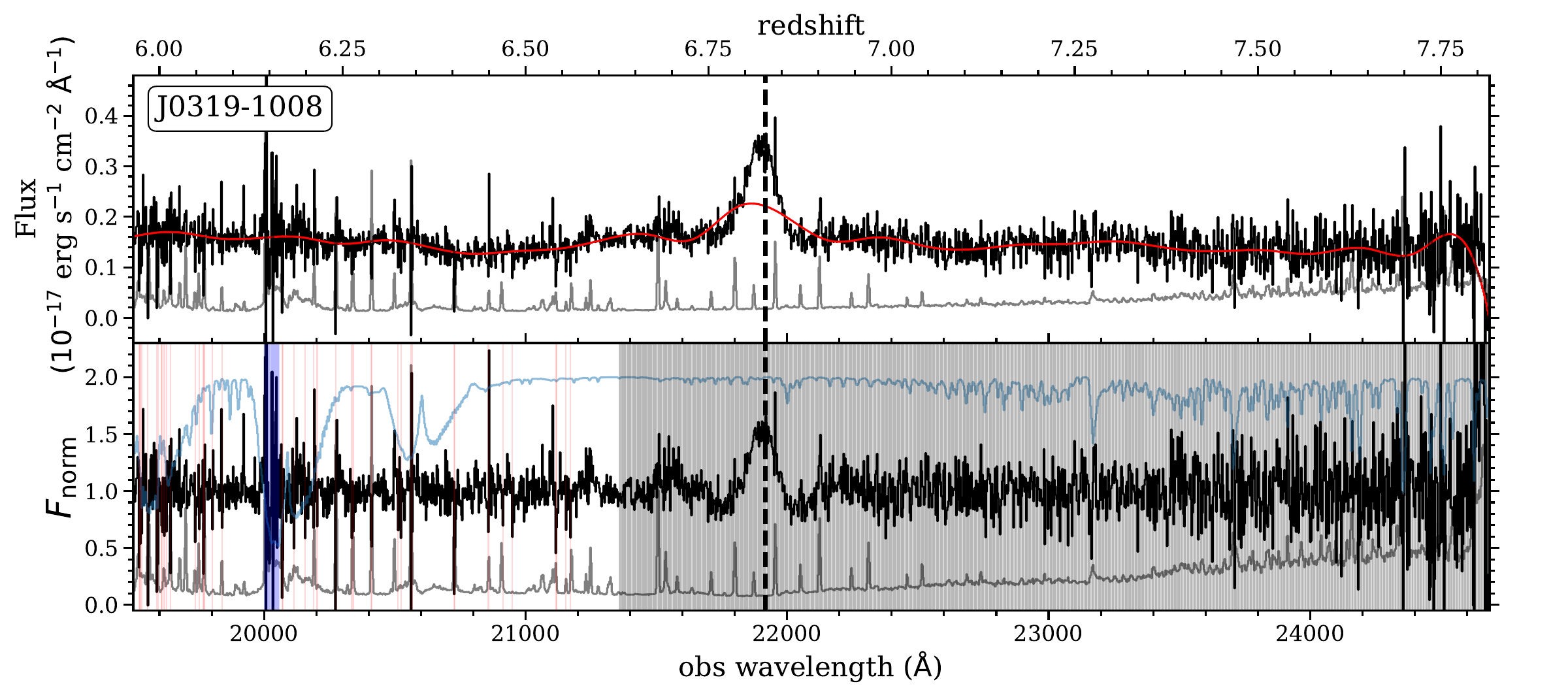}
\caption{Same as Figure \ref{specj0313}, but for J0411$-$0907 (top) and J0319$-$1008 (bottom)}.
\label{specj0411}
\end{figure*}

We fit for the quasar continuum using a B-spline 
with a breakpoint at every-60$^{\rm{th}}$ pixel (amounting to a breakpoint at every 2400 km/s spacing) and we mask 
strong absorbers identified by-eye before fitting the continuum. We use the same breakpoint spacing for all the quasars. Figures \ref{specj0313} to \ref{specj0411} 
show the spectra before and after continuum normalization for representative members of our dataset.
For each quasar, the gray shaded regions in the top panel indicate regions that are masked before continuum fitting due to
visually-detected strong absorbers and masks provided by \texttt{PypeIt} from the data reduction. In the bottom panels, we mask regions due to proximity to quasars and where $z > z_{\rm{QSO}}$ (gray), low SNR due to telluric absorption (blue), and detected CGM absorbers (red; see \S 4); these are excluded from the correlation function analyses.

We also mask pixels that fall within a quasar proximity zone, as these pixels will be affected by the quasar own ionizing radiation and produce biased results on the IGM neutral fraction. For the proximity zone (PZ) mask, we remove pixels within rest-frame $\Delta v = 7643$ km/s blueward from the quasar \ion{Mg}{II} emission line, where we define the \ion{Mg}{II} emission line to be the midpoint of the \ion{Mg}{II} $\lambda2796$\r{A} and $\lambda2804$\r{A} doublet at 2800 $\rm{\AA}$. Our choice of $\Delta v = 7643$ km/s follows \cite{Bosman2021}, who consider pixels at $<$ 1185 $\rm{\AA}$ for their Ly$\alpha$ optical depth study, which results in a proximity zone size of 7643 km/s, or 8.64 proper Mpc at the mean redshift of our quasars, which is $z$ = 7.16. We use a fixed PZ size for all quasars as there are indications that quasar PZ sizes do not evolve strongly with redshift \citep{Eilers2017,Satyavolu2023}, and even in the presence of evolution, our adopted value of 8.64 proper Mpc is rather conservative given that the typical PZ size for $z \sim 6$ quasars ranges from $2-7$ proper Mpc \citep{Eilers2017,Eilers2020,Satyavolu2023}.
The unshaded regions are used for the correlation function measurements, and they range from $z_{\rm{\ion{Mg}{II}}} = 5.96$ to $z_{\rm{\ion{Mg}{II}}} = 7.42$, with $z_{\rm{\ion{Mg}{II}, med}} = 6.469$, where $z_{\rm \ion{Mg}{II}} \equiv (\lambda_{\rm{obs}}\,/\, 2800 \, \rm{\AA}) - 1$ is the pixel redshift and $z_{\rm{\ion{Mg}{II}, med}}$ is the median value. Table \ref{tab:qso} includes the pathlength of each quasar that contributes towards the correlation function measurement, where the total pathlength of our dataset is $\Delta z = 9.75$ or $\Delta X = 47.1$. $\Delta X$ is the absorption pathlength, where $X(z) = \int_0^z (1 + z')^2 H_0/H(z')\, dz'$ \citep{BahcallPeebles1969}. We compute $\Delta X_{\rm{qso}} = X(z_i) - X(z_f) $ for each quasar and take the sum as the total absorption pathlength quoted above.

\begin{figure*}
\centering
\includegraphics[width=0.95\textwidth]{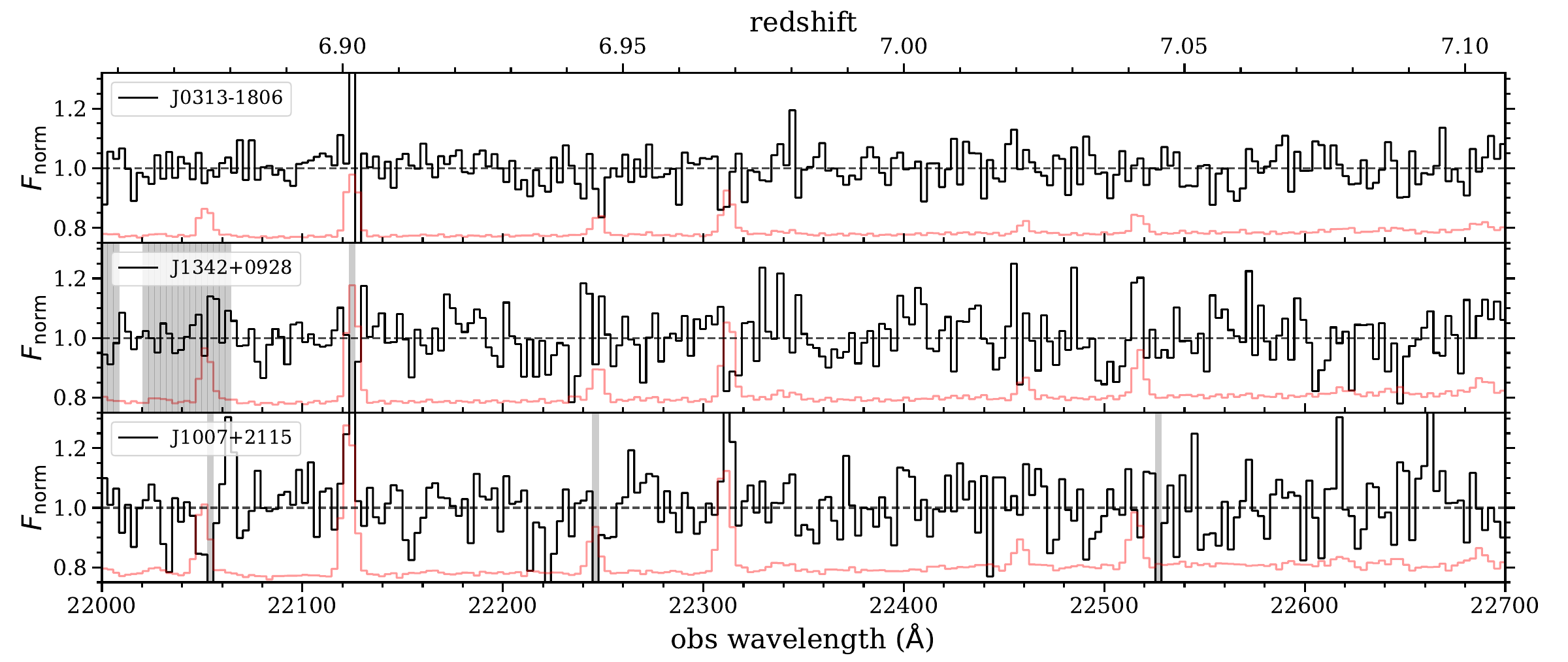}
\includegraphics[width=0.95\textwidth]{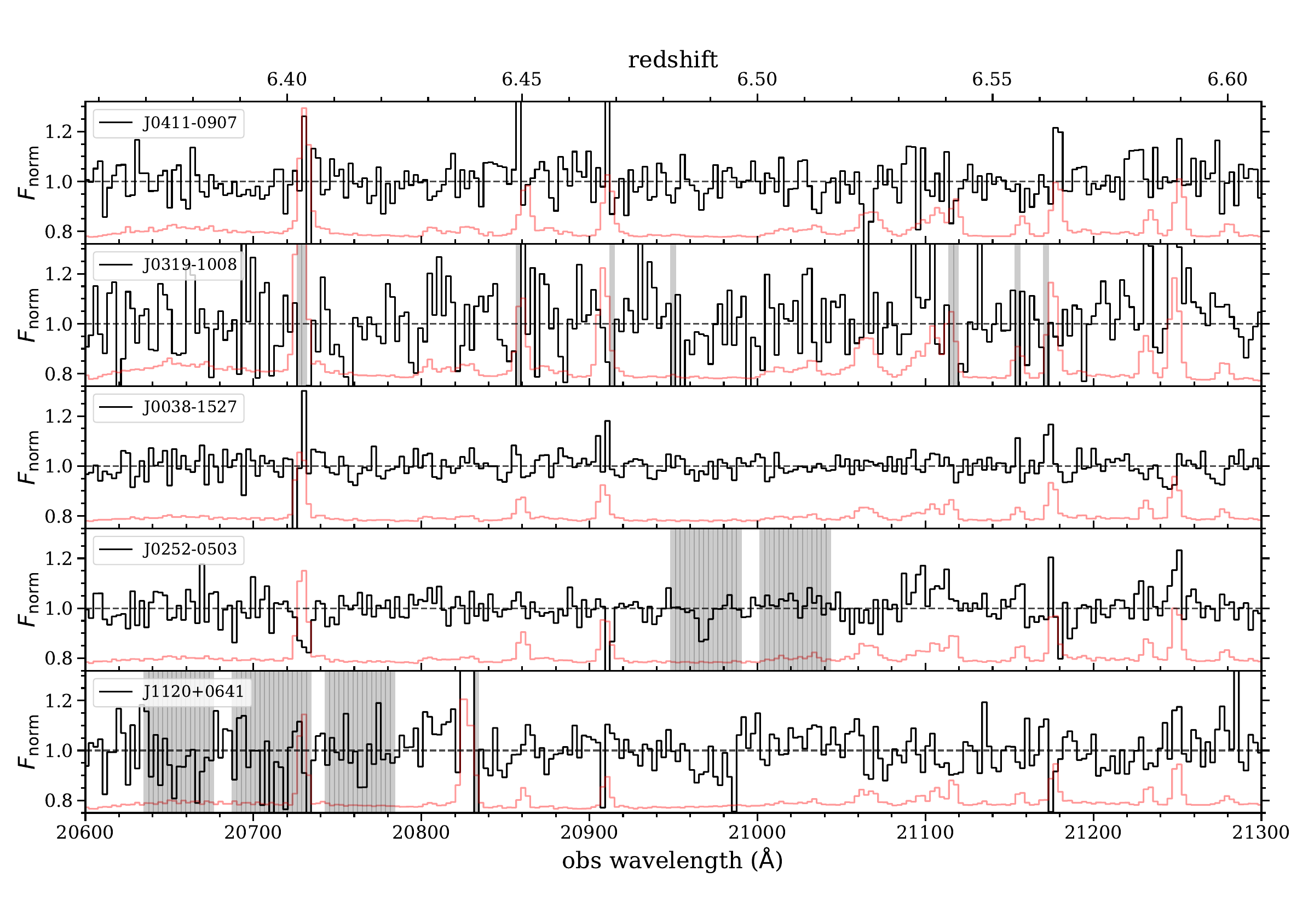}
\caption{Zooming in on regions of the \ion{Mg}{II} forest, where the top three panels are for the $z > 7.5$ quasars and the rest for the $z \leq 7.5$ quasars in our dataset. Red lines are the normalized noise, which are offset by some amount from zero for visualization purpose, while the shaded regions are masked (see \S \ref{observations}), although here we do not color code them like we did for Figure \ref{specj0313} - Figure \ref{specj0411}.}
\label{spec-mg2forest}
\end{figure*}

We discover that the noise in the quasar spectra produced by \texttt{PypeIt} are slightly over- and under-estimated and not normally distributed, and we apply correction factors to correct for this such that the noise distribution is Gaussian. To determine the correction factors, we compute the residual $\frac{1 - F_{\rm{norm}}}{\sigma_{\rm{norm}}}$ after applying all masks (including CGM masks described in \S \ref{masking}) and obtain the median absolute deviation (MAD) of the residual distribution for each quasar as the correction factor to the noise. We obtain correction factors of 0.972, 0.956, 0.908, 1.059, 1.086, 1.055, 0.93, 0.898, 0.88, 1.051 to the noise vectors of quasars no. 1 to no. 10 (see Table \ref{tab:qso}), respectively. After applying these corrections, the median SNR of our dataset is 13.7.
Figure \ref{spec-mg2forest} zooms in on certain regions of the \ion{Mg}{II} forest from the normalized spectra, where the typical flux fluctuations given our data quality is $\sim$ 10\%, or higher towards the redder part of the spectrum. 
Typical IGM absorbers are expected to produce percent-level fluctuations (see Figure 2 of \citetalias{Hennawi2021}), so attempting to detect them at this SNR level will be challenging.


\section{Masking \mgii\ absorbers of the CGM}
\label{masking}
As we are interested in measuring properties of the IGM, \ion{Mg}{II}\ absorbers within the virial radii of galaxies, namely those in their circumgalactic media (CGM), can bias our measurement (Figure 15 of \citetalias{Hennawi2021}). While the strongest absorbers can be identified by eye, weaker but more abundant
absorbers are difficult to detect visually. We briefly describe our procedures to mask out the CGM absorbers here, but see \citetalias{Hennawi2021} or \cite{Tie2022} for more details. 

\begin{figure*}
\centering
\includegraphics[width=\columnwidth]{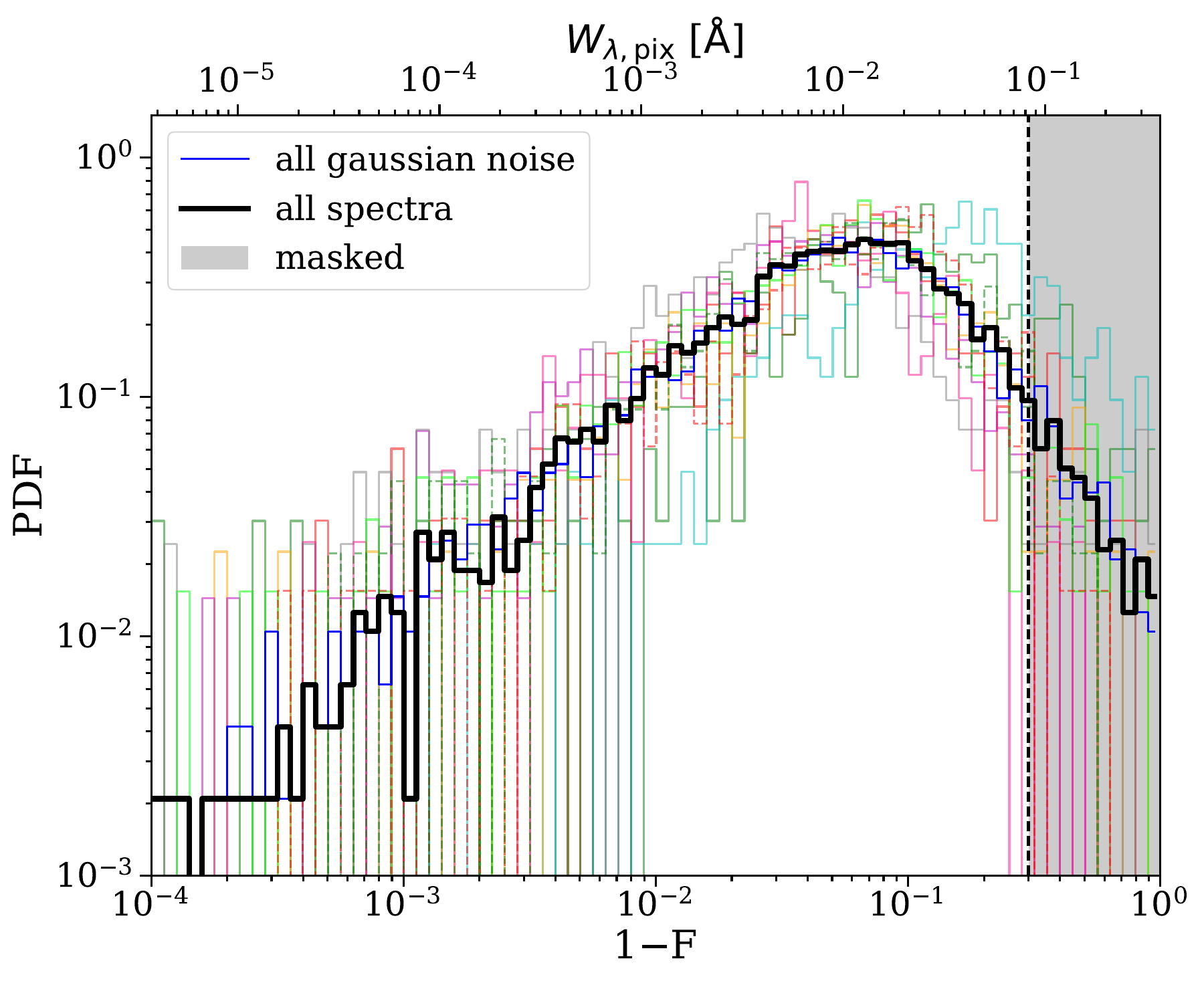}
\includegraphics[width=\columnwidth]{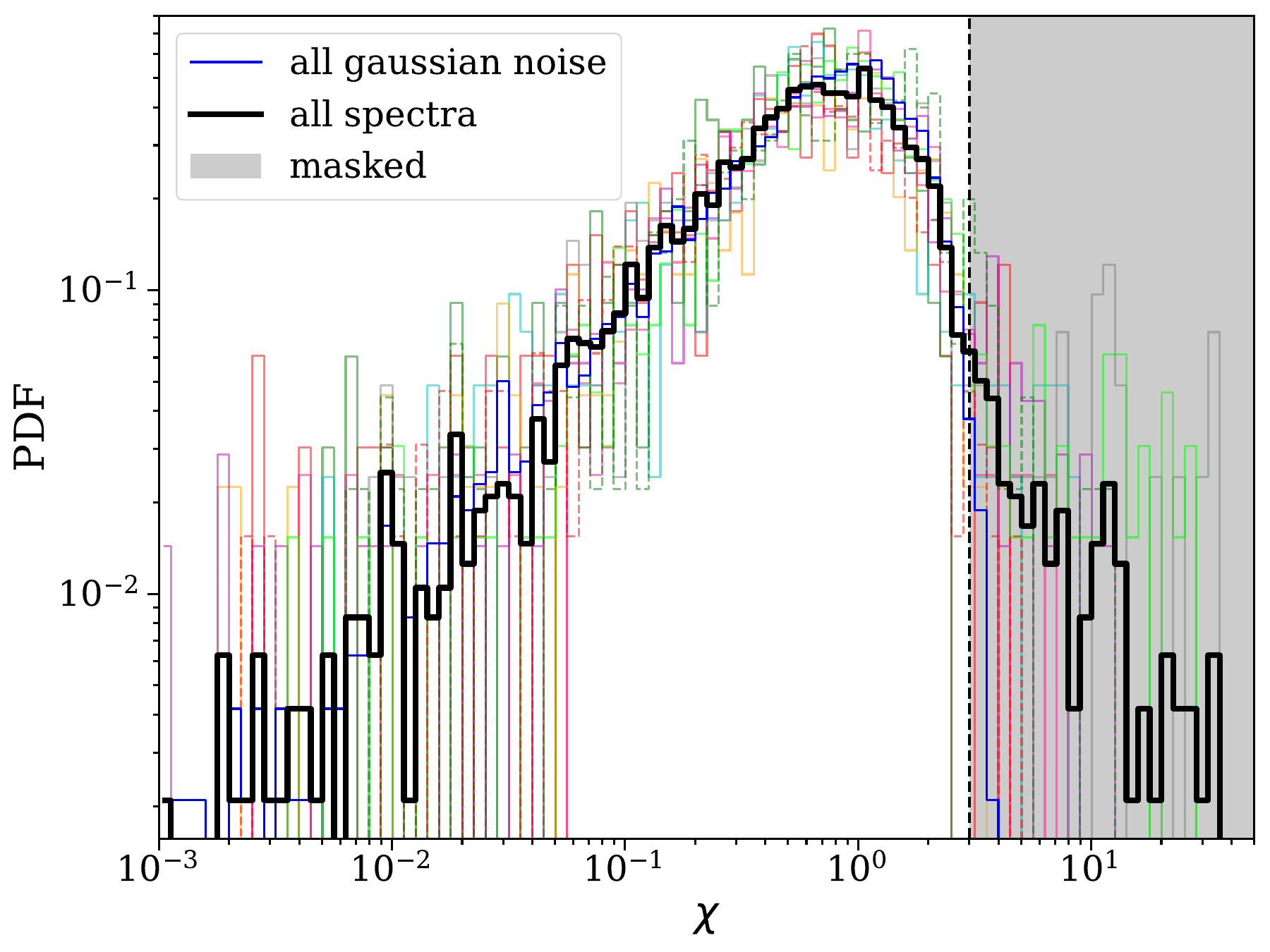}
\caption{Masking schemes to remove CGM absorbers based on the PDF of the flux decrement (left) and significance (right). The colored lines are the distributions for each quasar and the black lines are the distributions from all quasars. The dark blue lines are the distributions of Gaussian noise drawn from the corrected noise array from all quasars (see \S \ref{observations}). To remove CGM absorbers, we mask all pixels with $1 - \rm{F} > 0.3$ (shaded region in left panel), as well as those with $\chi > 3$ (shaded region in right panel) and $\pm 300$ km/s around each masked pixel.}
\label{flux-chi-pdf}
\end{figure*}

The first masking procedure uses the flux decrement ($1 - F$) probability distribution function to filter out CGM absorbers. The rationale behind this procedure is that the $1 - F$ PDF of the \ion{Mg}{II} forest will exhibit a tail towards large values (strong absorptions), which is sourced by rare strong CGM absorbers, and a peak at smaller value due to IGM absorbers (Figure 14 of \citetalias{Hennawi2021}).
Figure \ref{flux-chi-pdf} (left panel) shows the flux PDF for our data, where we mask all pixels with  $1 - F > 0.3$ (shaded region). The top axis shows the the pixel equivalent width, $W_{\lambda,\mathrm{pix}} \equiv (1-F)\,\Delta \lambda$, where $\Delta \lambda$ is the spectral pixel width at rest-frame \ion{Mg}{II} 2796 $\mathrm{\AA}$, which depends on the spectral sampling. Using our fixed common grid with $dv = 40$ km/s, $\Delta \lambda \sim 0.37 \,\rm{\AA}$.


While a flux cut helps remove strongly-absorbed pixels, it is usually insufficient for removing all affected pixels due to the broad absorption wings of CGM absorbers. As such, we supplement the flux cut with a second masking procedure, which identifies absorbers using the transmission field convolved with the profile of the \ion{Mg}{II}\ doublet. The convolved transmission field is known as the significance or $\chi$ field, 
\begin{align}
\chi(v) = \frac{\int [1-F(v')]W(|v - v'|) dv'}{\sqrt{\sigma_F^2(v')W^2(|v - v'|) dv'}},
\end{align}
where $\sigma_F^2$ is the flux variance (noise vector) of the data
and $W(v) = 1 - e^{-\tau(v)}$ is a matched filter with
$\tau(v)$ being the optical depth of a \ion{Mg}{II}\ doublet with a Gaussian velocity distribution, assuming $N_\mathrm{\mgii} = 10^{13.5}$ cm$^{-2}$ and Doppler parameter $b = \sqrt{2} \times \mathrm{FWHM}/2.35$, in which we use the FWHM of the respective quasar spectrum, except for the XSHOOTER spectrum of J1120$+$0641. For J1120$+$0641, we use a larger FWHM of 150 km/s, considering that we rebin all spectra to a fixed 40 km/s grid.
Note that $F$ is the continuum-normalized flux and $\sigma_F^2$ is the continuum-normalized variance. Given the $\chi$ spectrum, absorbers are identified by their extreme $\chi$ value. We mask out pixels above $\chi > 3$ as well as pixels $\pm 300$ km/s around each masked pixel to account for their absorption wings. Figure \ref{flux-chi-pdf} (right panel) shows the $\chi$ distribution of the data. In both panels, in addition to the distribution of each quasar (colored lines), we also plot the distribution for all quasars (black) and pure Gaussian noise sampled from the noise vectors of all quasars (blue). While the $1-F$ PDF from all quasars is similar to that of noise, the $\chi$ PDF shows evidence for fluctuations that are not pure noise at $\chi > 3$, since the noise PDF drops off at $\chi \sim 3$. Both PDFs also peak at some characteristic values as expected (see Figure 13 of \citealp{Hennawi2021}), although the $1-F$ PDF does not flatten off at large values and the plateau at large $\chi$ is weak in the $\chi$ PDF.
In summary, we mask 1) pixels with  $1 - \rm{F} > 0.3$ and 2) pixels with $\chi > 3$ and those $\pm 300$ km/s around each masked pixel.



\begin{figure*}
\includegraphics[width=\textwidth, height=0.41\textwidth,keepaspectratio]{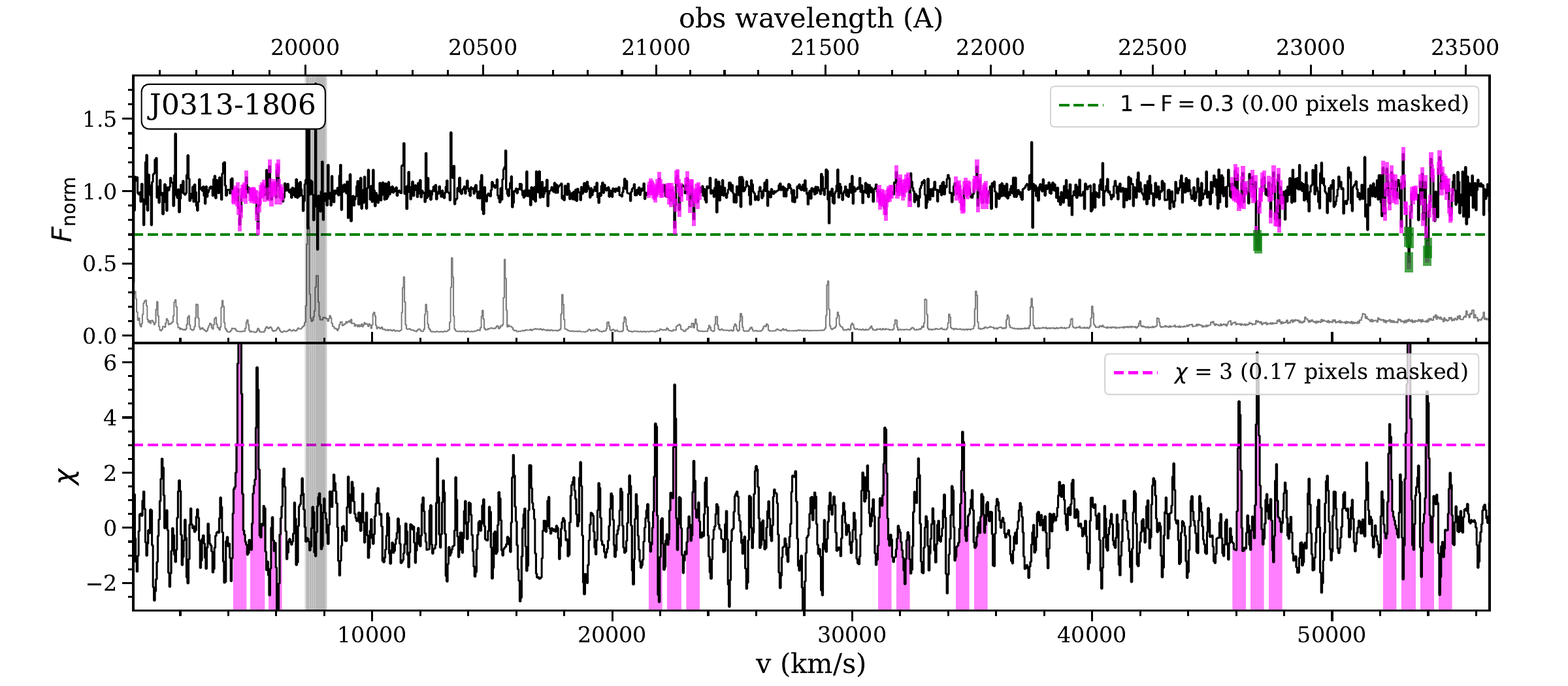}
\includegraphics[width=\textwidth, height=0.41\textwidth,keepaspectratio]{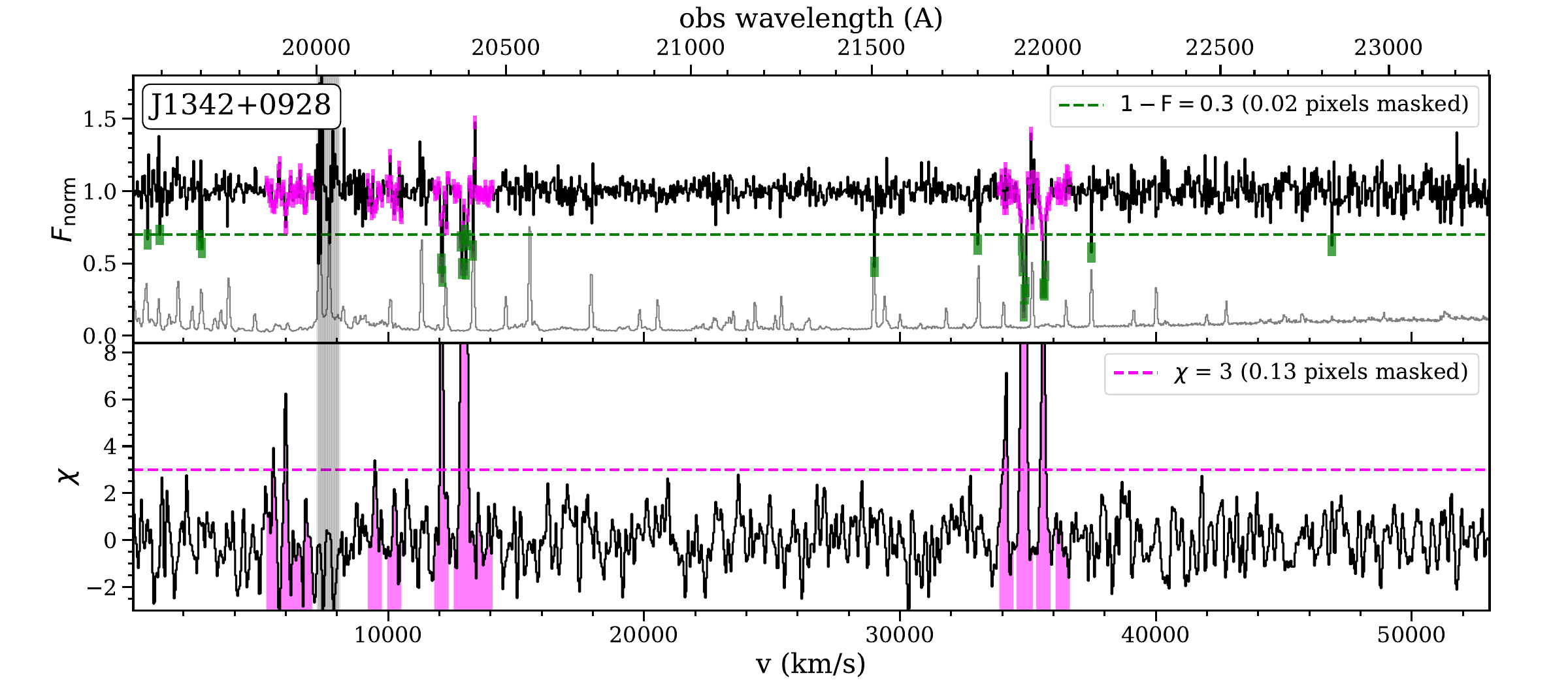}
\includegraphics[width=\textwidth, height=0.41\textwidth,keepaspectratio]{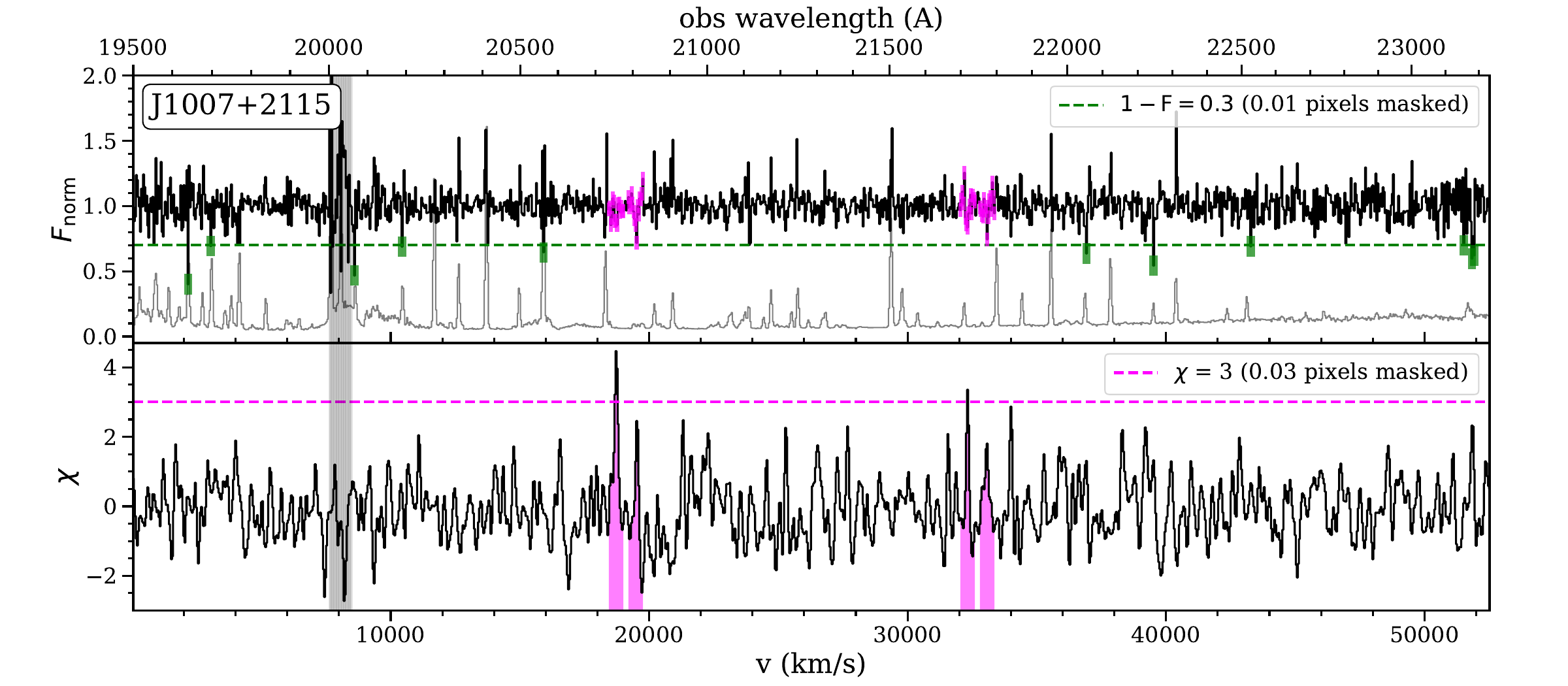}
\caption{Identified and subsequently masked CGM absorbers from J0313$-$1806 (top), J1342$+$0928 (middle), and J1007$+$2115 (bottom), where we have cut off regions of the spectra within the quasar proximity zone and beyond the quasar redshift. For each quasar, the top panel shows the continuum-normalized spectrum and the bottom panel shows the $\chi$ significance spectrum. The green and magenta horizontal lines indicate the thresholds for the $1 - \rm{F}$ and $\chi$ cuts, respectively (note that we additionally mask pixels $\pm 300$ km/s around pixels with $\chi > 3$). Pixels that are masked by the respective cuts are colored similarly. Gray shaded regions are regions of telluric absorptions. Our masking procedure results in a number of weak and strong absorber candidates, while recovering known strong absorbers.}
\label{masked-all-qso}
\end{figure*}

Figure \ref{masked-all-qso} shows the masked absorbers for the three highest-$z$ quasars in our dataset (we show the rest in Figure \ref{masked-allother-qso} and \ref{masked-allother-qso2} in Appendix B for brevity).
The green and magenta horizontal lines indicate our flux and $\chi$ cutoffs, respectively. We discover new strong absorber candidates in J0313$-$1806, J0038$-$1527, and newqso1, and recover known absorbers at $z=6.84$ \citep{Simcoe2020} and $z=6.27$ in the spectrum of J1342$+$0928. We recover two of the three weak absorbers identified by \cite{Bosman2017} in J1120$+$0641, and for the one that we do not detect, its chi values are $\sim 2$ and it is masked manually. These absorbers are indicated by the blue ticks in Figure \ref{masked-allother-qso} in Appendix B. 

\section{Results}
\label{results}

\subsection{Auto-correlation measurement of the \ion{Mg}{\uppercase{II}} forest}
\label{measurement}
We perform the correlation function measurement over three redshift bins:
\begin{enumerate}
  \item $5.96 \leq z_{\rm \ion{Mg}{II}} < 6.469$ (``low-$z$'' bin, where $z_{\rm{\ion{Mg}{II}, med}} = 6.235$)
  \item $6.469 \leq z_{\rm \ion{Mg}{II}} \leq 7.421$ (``high-$z$'' bin, where $z_{\rm{\ion{Mg}{II}, med}} = 6.715$)
  \item $5.96 \leq z_{\rm \ion{Mg}{II}} \leq 7.421$ (``all-$z$'' bin, with $z_{\rm{\ion{Mg}{II}, med}} = 6.469$)
\end{enumerate}
Recall that $z_{\rm \ion{Mg}{II}} \equiv (\lambda_{\rm{obs}}\,/\, 2800 \, \rm{\AA}) - 1$ is the redshift of the pixels that contributed to the correlation function and $z_{\rm{\ion{Mg}{II}, med}}$ is the median value. We first compute the relative flux fluctuation for each continuum-normalized quasar spectrum ($F$) as  
\begin{align}
    \delta_f \equiv \frac{F - \bar{F}}{\bar{F}}, 
\end{align}
where $\bar{F}$ is the mean flux of the entire dataset in each redshift bin. We compute $\bar{F}$ as the weighted average of the normalized flux for the entire dataset using the inverse variance of the corrected measurement noise (see below). 
We use the sub-optimal unbiased estimator for the correlation function as the weighted averaged of all pixel pairs $i,j$ \citep{Slosar2011}
\begin{align}
    \xi(\Delta v) = \frac{\sum_{i,j}w_{i}w_{j}\delta_{i}{\delta_{j}}}{\sum_{i,j} w_{i}w_{j}},
\label{eqn:cfone}
\end{align} 
where $\sum_{i,j}$ means summing over pixel pairs $i,j$ separated by the velocity lag $\Delta v$. Rewriting the sum as a sum over all pixel pairs from each quasar $k$, the correlation function can also be computed as the weighted average over the individual quasar correlation functions
\begin{align}
    \xi(\Delta v) &= \frac{\sum_k \sum_{i,j}w_{ik}w_{jk}\delta_{ik}{\delta_{jk}}}{\sum_k \sum_{i,j} w_{ik}w_{jk}} \nonumber \\
    &= \frac{\sum_k u_k \,\xi_k(\Delta v)}{\sum_k u_k} \nonumber \\
    &= \sum_k u_k^{\rm{eff}} \,\xi_k(\Delta v)
\label{eqn:cftwo}
\end{align}
where $u_k \equiv \sum_{i,j} w_{ik}w_{jk}$, $u_k^{\rm{eff}} \equiv \frac{u_k}{\sum_k u_k}$
is the result of normalizing $u_k$ (i.e. $\sum_k u_k^{\rm{eff}} = 1$), and  $\xi_k(\Delta v) = \sum_{i,j}w_{ik}w_{jk}\delta_{ik}\delta_{jk}/u_k$ is the weighted estimate of the correlation function of each quasar $k$. 

Along similar lines, $\bar{F}$ is computed as a weighted average via 
\begin{equation}
\bar{F} = \frac{\sum_{ki} w_{ik} F_{ik}}{\sum_{ki}w_{ik}},
\end{equation}
where $F_{ik}$ is the continuum normalized flux of the $i$th spectral pixel in the $k$th quasar.  The values of $\bar{F}$ are 1.0015, 1.001, 1.0018 for the all-$z$, high-$z$, and low-$z$ bins, respectively, after masking out CGM absorbers. 

For the weights, We use the inverse variance of the corrected measurement noise, $w_{ik} = 1/\sigma^2_{ik}$, where 
$\sigma^2_{ik}$ is the error of the continuum normalized flux. While one should technically also include the variance due to the intrinsic fluctuations in the \ion{Mg}{II} forest, we neglect that term here as we are noise-dominated --- based on \citetalias{Hennawi2021}, for an IGM model with [Mg/H] = $-3.50$ ($-4.50$), the typical fluctuation in the \ion{Mg}{II} forest is $\sim 5\%$ (2\%), which requires a SNR of at least 20 (50), whereas the median SNR of our dataset is 13.7 (considering only the top five highest SNR quasars, the mean SNR is still $\sim$ 18). 






We measure the correlation function of the \ion{Mg}{II} forest from $\Delta v_\mathrm{min} = 10$ km/s to $\Delta v_\mathrm{max} = 3500$ km/s. Given that the correlation function is expected to rise at small scales and peak at the doublet separation (768 km/s), and flatten at large velocity separations, a fine bin size is needed to capture the small-scale rise and the correlation function peak, whereas a coarse bin size is sufficient at large scales. Therefore, to save computation time during the forward modeling step (\S \ref{constraints}), we adopt a heterogenous binning where we use a bin size of 80 km/s at $\Delta v < 1500$ km/s and a bin size of 200 km/s at $\Delta v \geq 1500$ km/s. Since we are using the weighted average over all quasars as the measurement for the dataset, given the heteregeneous SNR and pathlength of the quasars, they do not contribute equally to the final measurement. Table \ref{tab:qso} shows the effective weights, $u_k^{\rm{eff}}$, which quantifies the contribution of each quasar to the total correlation function for each redshift bin. For the all-$z$ bin, J0313$-$1806 has the highest weight (33\%), followed by J0038$-$1527 (22\%), J1342+0928 (14\%), and J0252$-$0503 (12\%). Similarly, the high-$z$ bin measurement is dominated by J0313$-$1806, J0038$-$1527, and J1342+0928, while the low-$z$ bin measurement is dominated by J0313$-$1806, J0038$-$1527, and J0252$-$0503. Given the unequal contributions of individual quasars to the correlation function measurement, we can compute a so-called effective sample size \citep{Kish1995}, which is an estimate of the sample size required for a random process where all data points carry uniform weight, to achieve the same level of variance as a random process with non-uniform weights. 
The effective sample size is given by $n_{\rm{eff}} = \frac{(\sum^i_{n} u^{\rm eff}_i)^2}{\sum^i_{n} (u^{\rm eff})^2}$, where $u^{\rm eff}_i$ is the weight of the $i$-th data point. We find values of $n_{\rm{eff}} =$ 4.9, 4.3, and 5.1 for the all-$z$, high-$z$, and low-$z$ bins, respectively, which is significantly less than the actual number of quasars in the dataset (ten).





\begin{figure*}
\centering
\includegraphics[width=\textwidth]{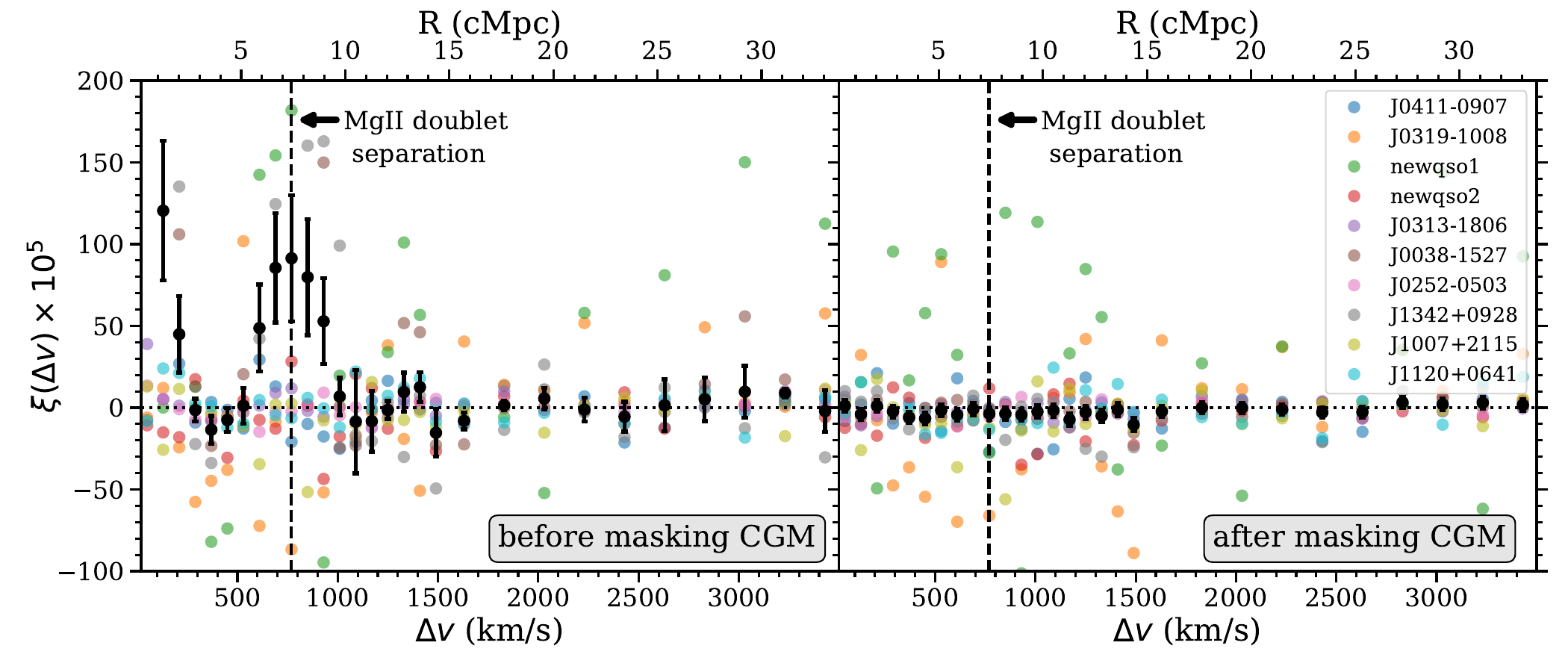}
\caption{The correlation function of the \ion{Mg}{II} forest from our quasar dataset measured in the ``all-$z$'' bin. The colored points are the measurements from each quasar and the black points are the average from all quasars. Note that we are using an unconventional binning where the bin size is 80 km/s at $\Delta v < 1500$ km/s and 200 km/s at $\Delta v \geq 1500$ km/s, hence the sparser data points at large scales. The top axes are computed at the median pixel redshift of this bin, $z_{\rm{\ion{Mg}{II}, med}} = 6.469$. \textit{Left:} The correlation function before we mask out CGM absorbers identified in \S \ref{masking}, where we see an obvious peak at the velocity separation of the \ion{Mg}{II} doublet (768 km/s; dashed vertical lines), as well as a small-scale rise due to the absorbers velocity structures. \textit{Right:} Results after masking out CGM absorbers, where the correlation function is consistent with noise.}



\label{mgii-cf-allz}
\end{figure*}

\begin{figure*}
\centering
\includegraphics[width=\textwidth]{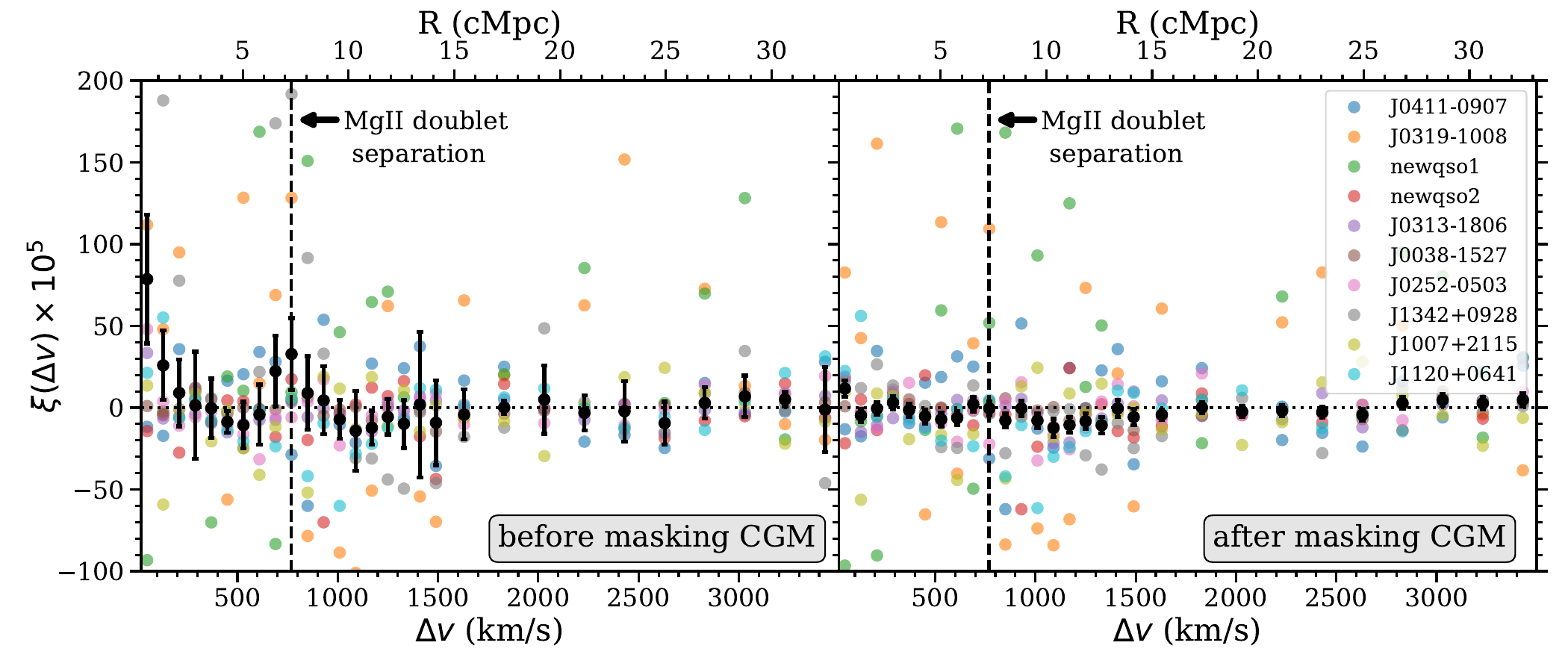}
\caption{Same as Figure \ref{mgii-cf-allz}, but showing the correlation function of the \ion{Mg}{II} forest from our quasar dataset measured in the ``high-$z$'' bin, where $z_{\rm{\ion{Mg}{II}, med}} = 6.715$.}
\label{mgii-cf-highz}
\end{figure*}

\begin{figure*}
\centering
\includegraphics[width=\textwidth]{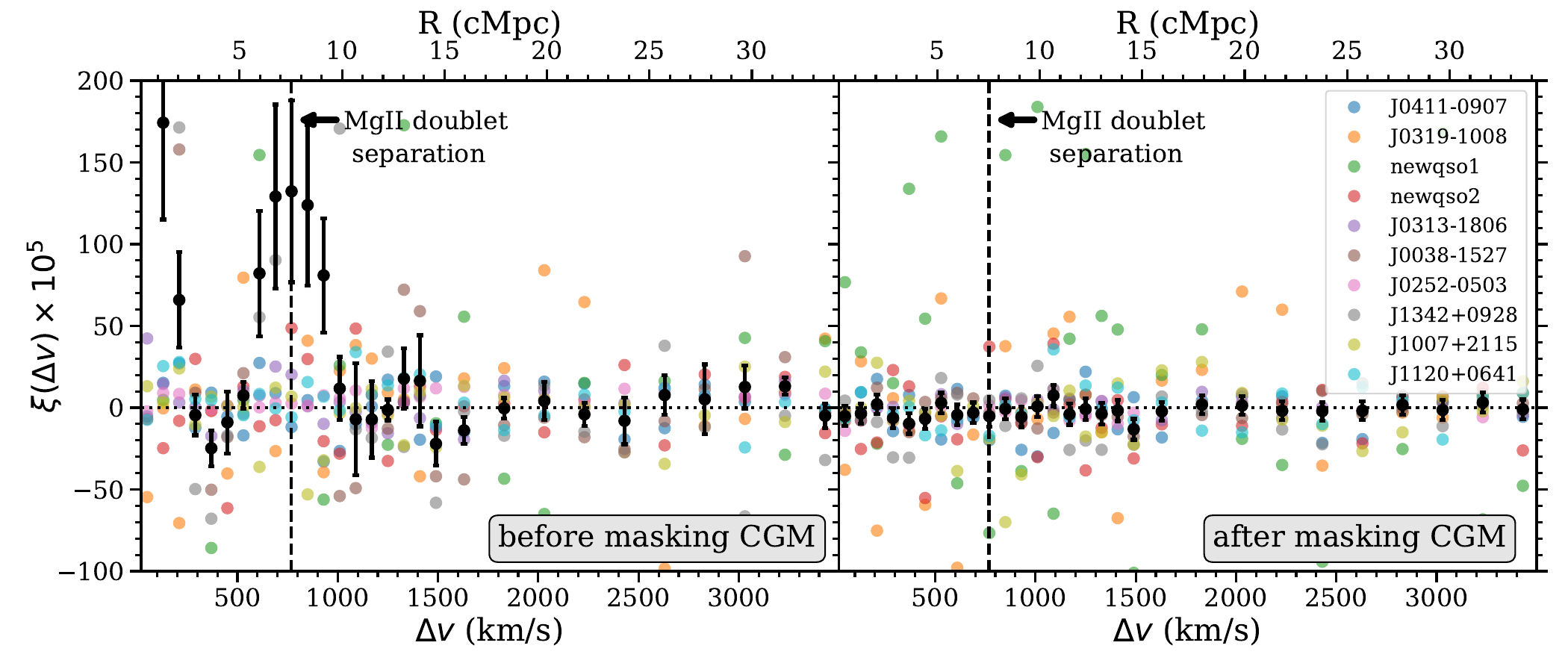}
\caption{Same as Figure \ref{mgii-cf-allz}, but showing the correlation function of the \ion{Mg}{II} forest from our quasar dataset measured in the ``low-$z$'' bin, where $z_{\rm{\ion{Mg}{II}, med}} = 6.235$.}
\label{mgii-cf-lowz}
\end{figure*}

Figures \ref{mgii-cf-allz} to \ref{mgii-cf-lowz} show the correlation functions before and after we filter out CGM absorbers (see \S \ref{masking}) in the three redshift bins. The colored points are the measurements for each quasar and the black points are the average from all quasars (Eqn \ref{eqn:cftwo}). In the left panels of the figures, we detect a strong peak in the correlation function at the \ion{Mg}{ii} doublet separation (768 km/s) that is due to strong CGM absorbers, as well as a rise at small velocity lags due to the small-scale velocity structure of the absorbers. The detection of the \ion{Mg}{ii} peak at the correct location provides a sanity check of our methodology. We roughly estimate the error bars on the correlation function in the left panels as the dispersion between the individual correlation functions divided by $\sqrt{N_q}$, with $N_q = 10$ being the total number of quasars in our dataset. We note that this error estimation is technically incorrect as our final correlation function measurement is a weighted one, but since we are only using these errors for visual purposes and given that bootstrapping a weighted measurement with a very small effective sample size ($\sim 5$) is challenging, we deem this to be sufficient. In the right panels, after masking out these strong absorbers, we detect a null signal consistent with noise in all redshift bins. Here, the error bars are obtained by taking the square root of the diagonal elements of the best-fit covariance matrix computed in \S 5.2.

\begin{figure}
\centering
\includegraphics[width=\columnwidth]{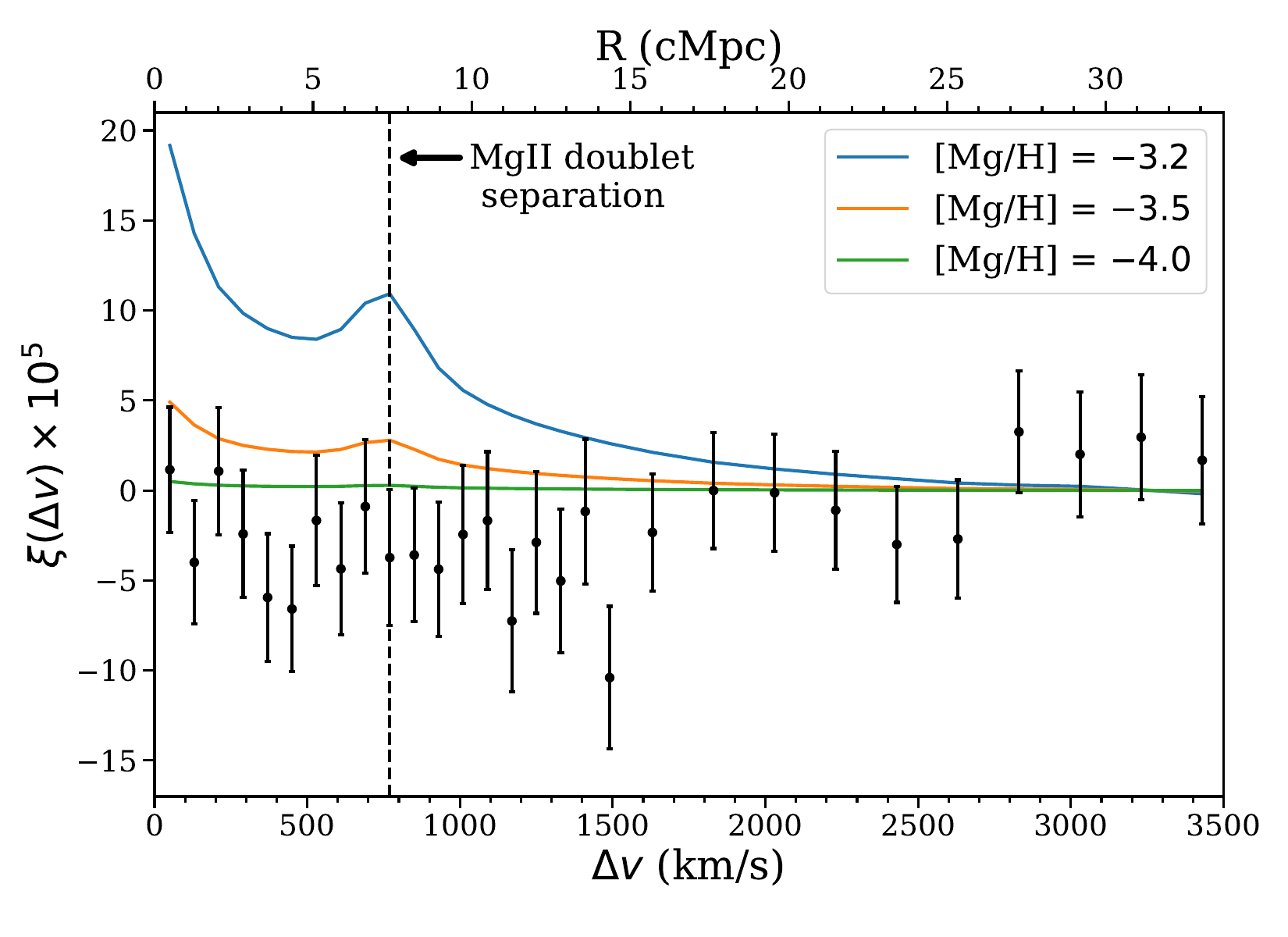}
\caption{Comparison of three IGM models from \citetalias{Hennawi2021} with varying [Mg/H] but fixed $x_{\rm{\ion{H}{I}}}$ = 0.5, indicated by the colored lines, with our correlation function measurement in the ``all-$z$'' bin after masking out CGM absorbers, which are the black points.}
\label{mgii-cf-compare-models}
\end{figure}

We qualitatively compare our measurements with several IGM models from \citetalias{Hennawi2021}, with varying [Mg/H] but fixed $x_{\rm{\ion{H}{I}}}$ = 0.5, assuming FWHM = 120 km/s and sampling of 3 to simulate our data grid but noiseless skewers otherwise, as shown in Figure \ref{mgii-cf-compare-models}, where by-eye we can rule out models with high [Mg/H] $> -3.5$. 
We perform a more rigorous inference in the next section. 


\subsection{Constraints on $x_{\rm{HI}}$ and [Mg/H] from Bayesian statistical inference}
\label{constraints}
We forward model our quasar dataset and perform Bayesian inference to place constraints (upper limits) on the IGM metallicity and neutral fraction. We create 1000 realizations of each quasar in our dataset using the IGM models from \citetalias{Hennawi2021} spanning from $x_{\rm{\ion{H}{I}}}$ = 0 to $x_{\rm{\ion{H}{I}}}$ = 1.0 in bins of 0.02 and from $[\mathrm{Mg/H}] = -5.5$ to $[\mathrm{Mg/H}] = -2.5$ also in bins of 0.02. These IGM models are produced by combining a $\texttt{Nyx}$ hydrodynamical simulation \citep{Almgren2013,Lukic2015} of the pre-reionized IGM at $z=7.5$, assuming the IGM is uniformly enriched with metals, with a modified version \citep{Davies2022} of the \texttt{21cmFAST} semi-numerical simulations \citep{Mesinger2011} of the reionization topology. The \texttt{Nyx} simulation was performed with a box size of 40 cMpc $h^{-1}$ on a 2048$^3$ grid and we analyze the output at a single redshit snapshot at $z=7.5$. The neutral fraction $x_{\rm{\ion{H}{I}}}$ from \texttt{21cmFAST} was computed on a 256$^3$ grid over the \texttt{Nyx} simulation domain and evaluated at $z=7.5$. Our \texttt{21cmFAST} simulations adopt a fixed ionizing mean free path of 20 cMpc and adjust the ionizing efficiency to obtain the full range of $x_{\rm{\ion{H}{I}}}$ models.


We forward model regions of the spectra outside of the quasar proximity zones. After convolving the mock \ion{Mg}{II} forest skewers
with the FWHM and sampling of the respective instrument that the quasar is observed with, 
we interpolate them onto the observed coarse wavelength grid with a pixel size of 40 km/s 
Finally, we randomly sample the corrected noise vectors assuming a Gaussian distribution and these nois realizations to the simulated spectra.
As the pathlength of a quasar in our dataset is longer than the pathlength of a mock skewer
($\sim$ 60,000 km/s vs. 6567 km/s, see \citetalias{Hennawi2021}), each forward-modeled spectrum is made up of $n$ randomly-drawn skewers such that the total pathlength of the two matches, where remaining pixels in the last skewer are not used in the analyses.

\begin{figure*}
\includegraphics[width=\textwidth]{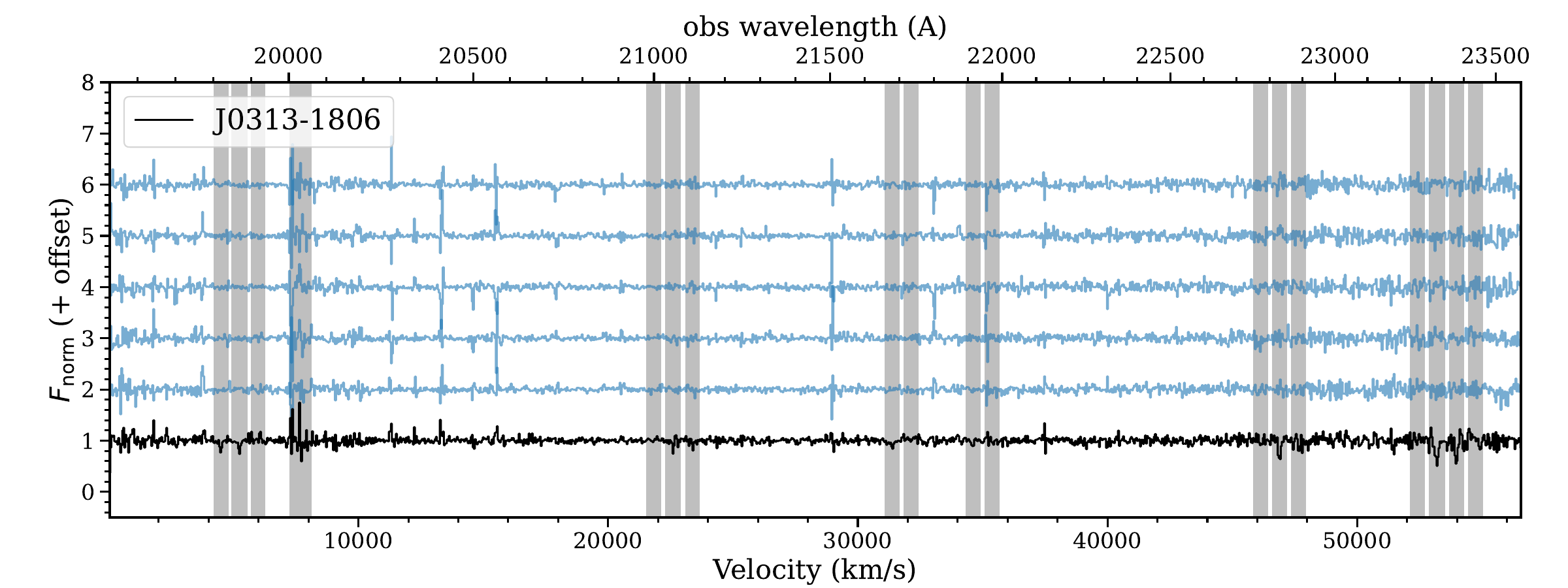}
\includegraphics[width=\textwidth]{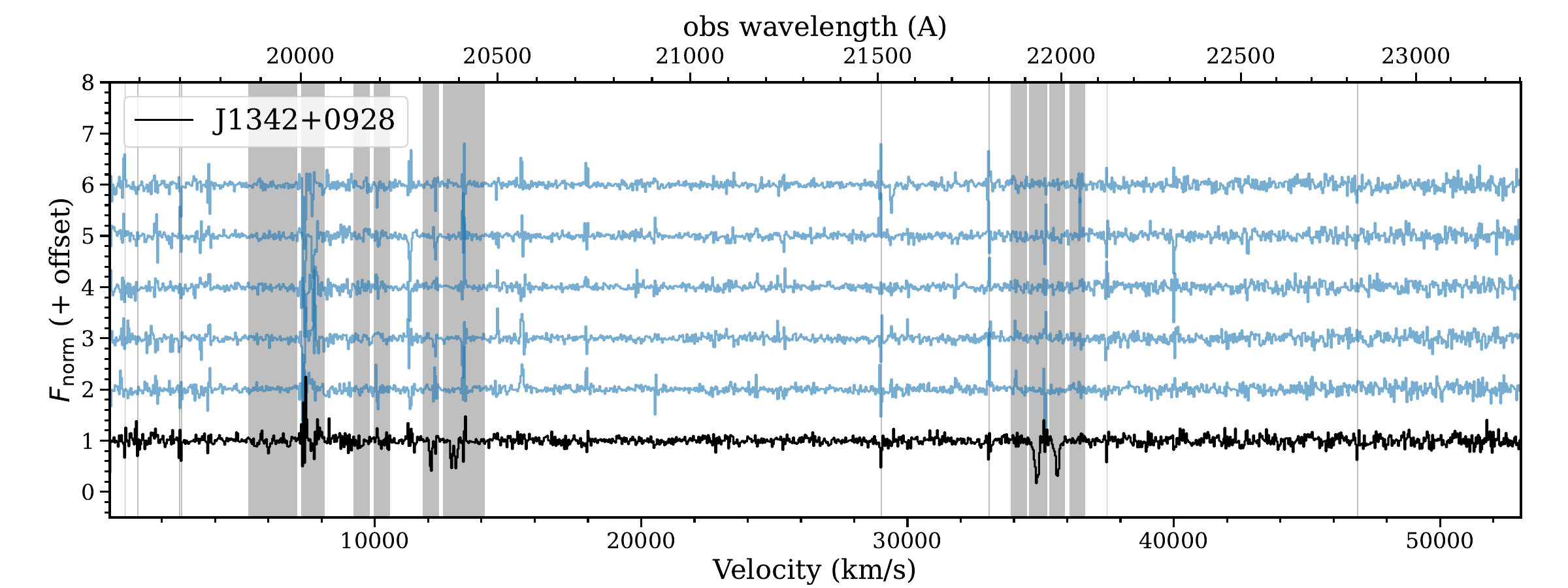}
\includegraphics[width=\textwidth]{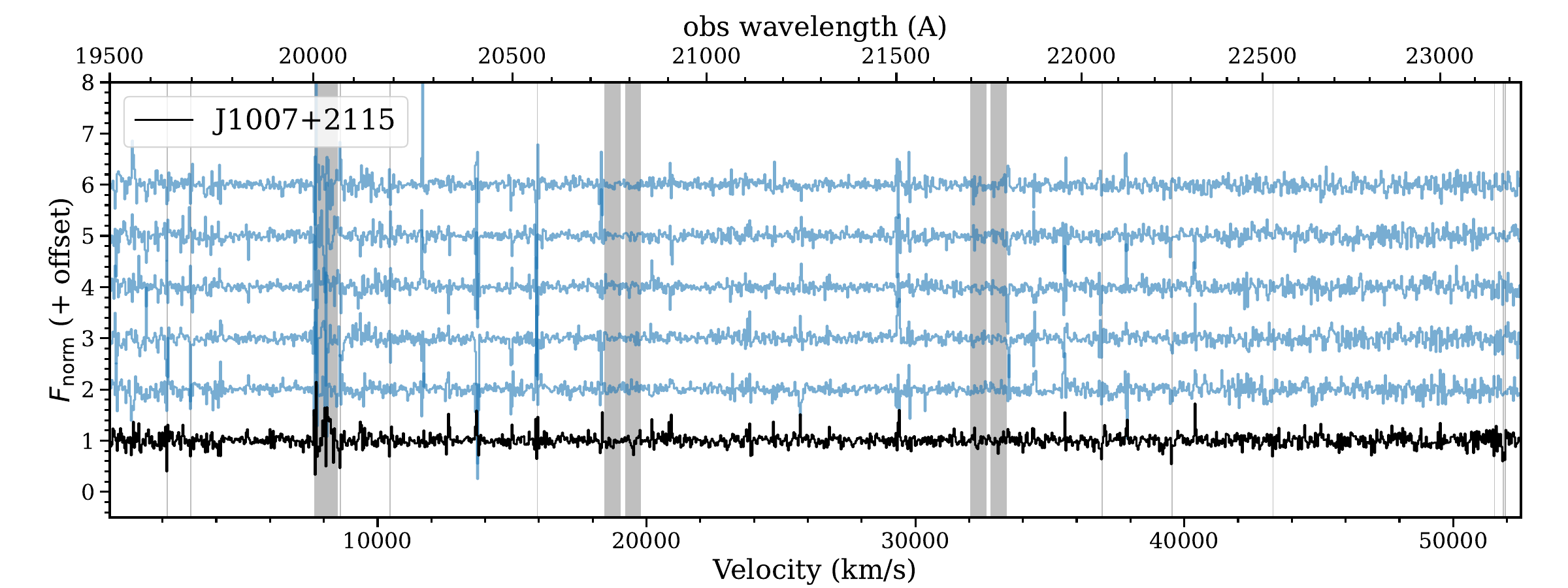}
\caption{Comparison of real spectrum (black) and examples of forward-modeled spectra (blue) for J0313$-$1806 (top), J1342$+$0928 (middle), and J1007$+$2115 (bottom), assuming an IGM model with (${x_{\rm{\ion{H}{I}}}, \rm{[Mg/H]}}) = (0.50, -4.50)$. Each forward-modeled spectrum is convolved with the respective FWHM and spectral sampling of the real data and the noise are randomly drawn from the data noise assuming a Gaussian distribution. The shaded regions are the masked regions from the real data, which includes masked CGM absorbers.}
\label{forwardmodels}
\end{figure*}

We generate $10^6$ forward-modeled datasets in total, where a mock dataset consists of randomly drawing one mock spectrum (i.e. multiple mock skewers per spectrum) from the 1000 realizations of each quasar.
Finally, we apply the same masks and weights as the data to the forward-modeled data when computing the mock correlation function. 
We repeat the forward modeling in all three redshift bins (low-$z$, high-$z$, and all-$z$). Figure \ref{forwardmodels} shows a comparison of real and forward-modeled spectra for the top three highest-$z$ quasars in our dataset (the rest are shown in Figure \ref{forwardmodels2} and \ref{forwardmodels3} in Appendix C), assuming an IGM model with (${x_{\rm{\ion{H}{I}}}, \rm{[Mg/H]}}) = (0.50, -4.50)$.

We perform Markov Chain Monte Carlo (MCMC) analysis using \texttt{EMCEE} \citep{Foreman-Mackey2013}, assuming a multivariate Gaussian likelihood as follows,
\begin{multline}
L(\hat{\xi}(\Delta v) \, | \, {x_{\rm{\ion{H}{I}}}, \rm{[Mg/H]}}) =
\\ \frac{1}{\sqrt{(2\pi)^k \mathrm{det}(\mathrm{\mathbf{C}})}}\mathrm{exp}(-\frac{1}{2}\mathbf{d}^T \mathbf{C^{-1}} \mathbf{d}), 
\end{multline}
where \textbf{C} is the covariance matrix, $\mathrm{det}(\mathrm{\mathbf{C}}$) is the determinant of the covariance matrix, $k = 29$ is the number of velocity bins over which we compute the correlation functions, and \textbf{d} = $\hat{\xi}(\Delta v) - \xi(\Delta v \, | \, {x_{\rm{\ion{H}{I}}}, \rm{[Mg/H]}})$ is the difference between the correlation function measured from the mock dataset $\hat{\xi}(\Delta v)$ and the model correlation function $\xi(\Delta v \, | \, {x_{\rm{\ion{H}{I}}}, \rm{[Mg/H]}})$ (henceforth abbreviated as $\xi(\Delta v)$). As a forward-modeled spectrum consists of $n$ random skewers, we first compute the correlation function of each skewer individually and take the weighted average over these skewers as the correlation function for each spectrum. The correlation function for the mock dataset, $\hat{\xi}(\Delta v)$, is computed as the weighted average over the individual quasar correlation function according to Eqn \ref{eqn:cftwo}. We compute the model correlation function $\xi(\Delta v)$ as the average over the individual noiseless correlation functions computed from the noiseless mock spectra. As discussed in \S \ref{measurement}, we measure the correlation function from a separation of 10 km/s to 3500 km/s in bins of 80 km/s at $\Delta v < 1500$ km/s and in bins of 200 km/s at $\Delta v \geq 1500$ km/s. 


While the covariance matrix can in principle be calculated by bootstrapping from the actual data, given our small dataset where the effective sample size is $\sim$5, the covariance matrix estimated from the data will be too noisy. As such, we compute the elements of the covariance matrix from the forward models via
\begin{align}
    C_{ij} 
    &= \langle [\hat{\xi}(\Delta v) - \xi(\Delta v)]_i [\hat{\xi}(\Delta v) - \xi(\Delta v)]_j \rangle,
\end{align}
where $i$ and $j$ refers to different bins of $\Delta v$ and the average is taken over the $10^6$ realizations of mock datasets. 

For our MCMC sampling, we assume a flat linear prior on $x_{\rm{\ion{H}{I}}}$ from 0 to 1 and a flat log prior on [Mg/H] from $-5.5$ to $-2.5$. To speed up the MCMC sampling, we interpolate the covariance matrix, its determinant, and the model correlation function $\xi(\Delta v)$ computed on our coarse model grid onto a finer grid with bins of 0.001 in $x_{\rm{\ion{H}{I}}}$ and bins of 0.003 in [Mg/H].

As we describe in detail in Appendix A, a challenge that we encountered was that the MCMC samples appeared to be visually inconsistent with the data points and error bars. We determined that this issue arises from the presence of correlated negative correlation function values at some velocity lags in the data, which do not occur in the forward models.  These strong correlated data points  are improbable given the covariance matrix and thus yield MCMC  samples and best-fit models that are visually inconsistent with the measurements. These spurious correlations arise from some unknown systematic in the data, which is most likely to be correlated noise resulting from sky-subtraction systematics.  We developed a procedure that allows us to mask specific problematic velocity lags, to ensure sensible inference, as we describe in detail in Appendix A. 

Figure \ref{allz-mcmc} shows the resulting parameter constraints in the all-$z$ bin.  Unsurprisingly, due to our null detection, we are unable to place any strong constraints on either parameters. To infer an upper limit on [Mg/H], we repeat the MCMC sampling with a flat linear prior on [Mg/H] and obtain an upper limit of $ -3.73$ at 95\% C.L. The reason to adopt a linear prior for the upper limit is because the resulting upper limit is strongly affected by the prior range for a log prior, but not so for a linear prior. 
Figure \ref{highz-mcmc} and \ref{lowz-mcmc} show the constraints on the high-$z$ and low-$z$ redshift bins, respectively. The corresponding 95\% C.L. upper limits are [Mg/H] $< -3.49$ and [Mg/H] $< -3.71$ for the high-$z$ and low-$z$ bins, respectively.

\begin{figure*}
\includegraphics[height=0.35\textheight, width=0.62\textwidth]{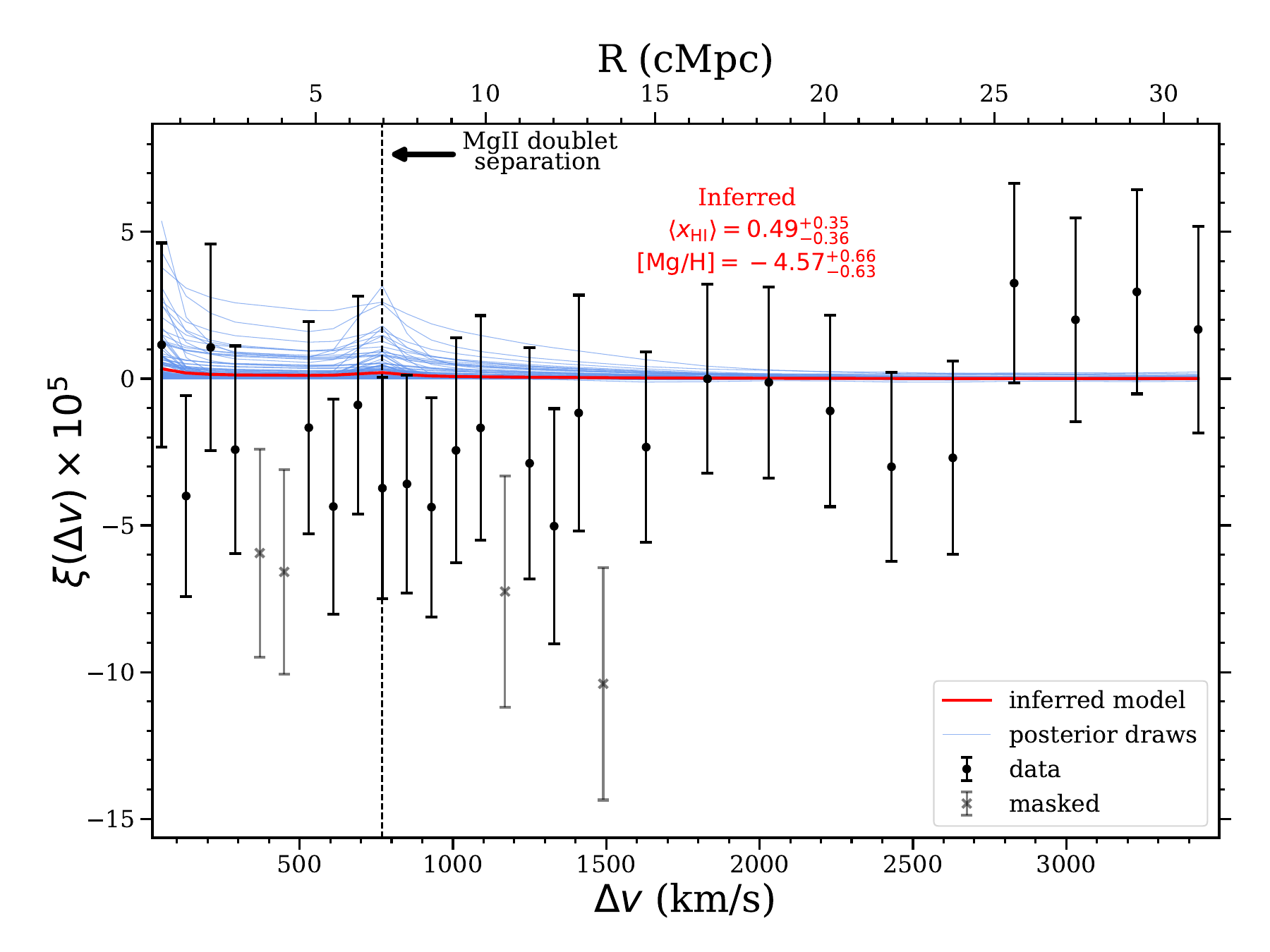}
\includegraphics[height=0.29\textheight, width=0.37\textwidth]{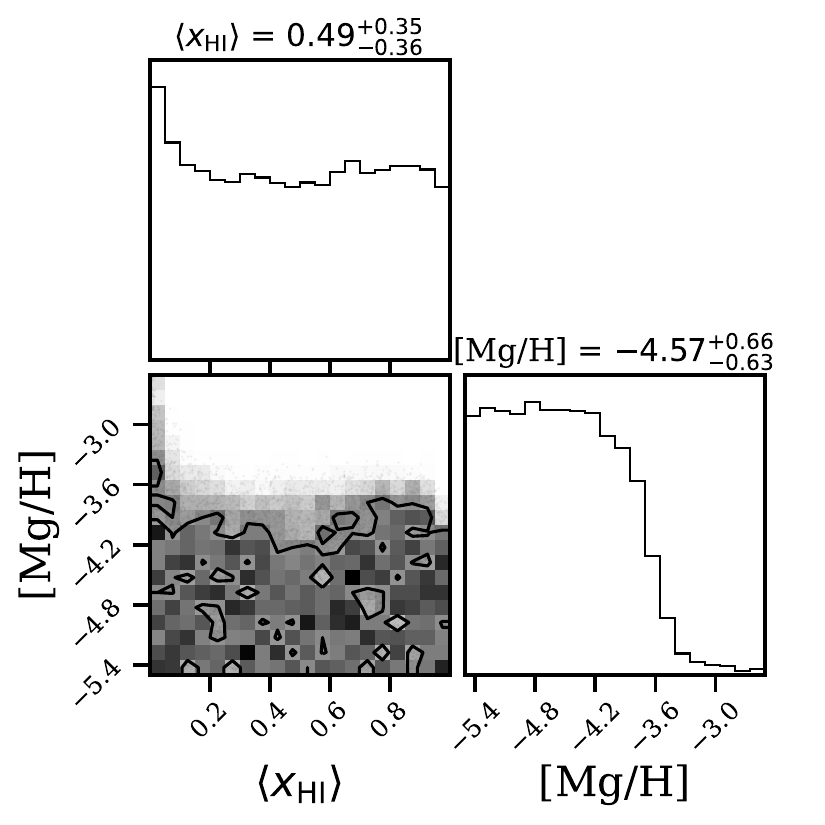}
\caption{Results from our MCMC analyses in the all-$z$ redshift bin ($z_{\rm{\ion{Mg}{II},med}} = 6.469$). \textit{Left:} Correlation function measured from our dataset compared with 200 random draws (blue lines) and the mean inferred model (red line) from the MCMC posterior distribution. The black points are measurements from our dataset, where we have masked the gray points before running the MCMC sampling (see Appendix A). The error bars are computed from the diagonal elements of the covariance matrix of the inferred model. \textit{Right:} Corner plot from MCMC sampling of the posterior distribution. The 95\% C.L. upper limit on [Mg/H] is $-3.73$.}
\label{allz-mcmc}\
\end{figure*}

\begin{figure*}
\includegraphics[height=0.35\textheight, width=0.62\textwidth]{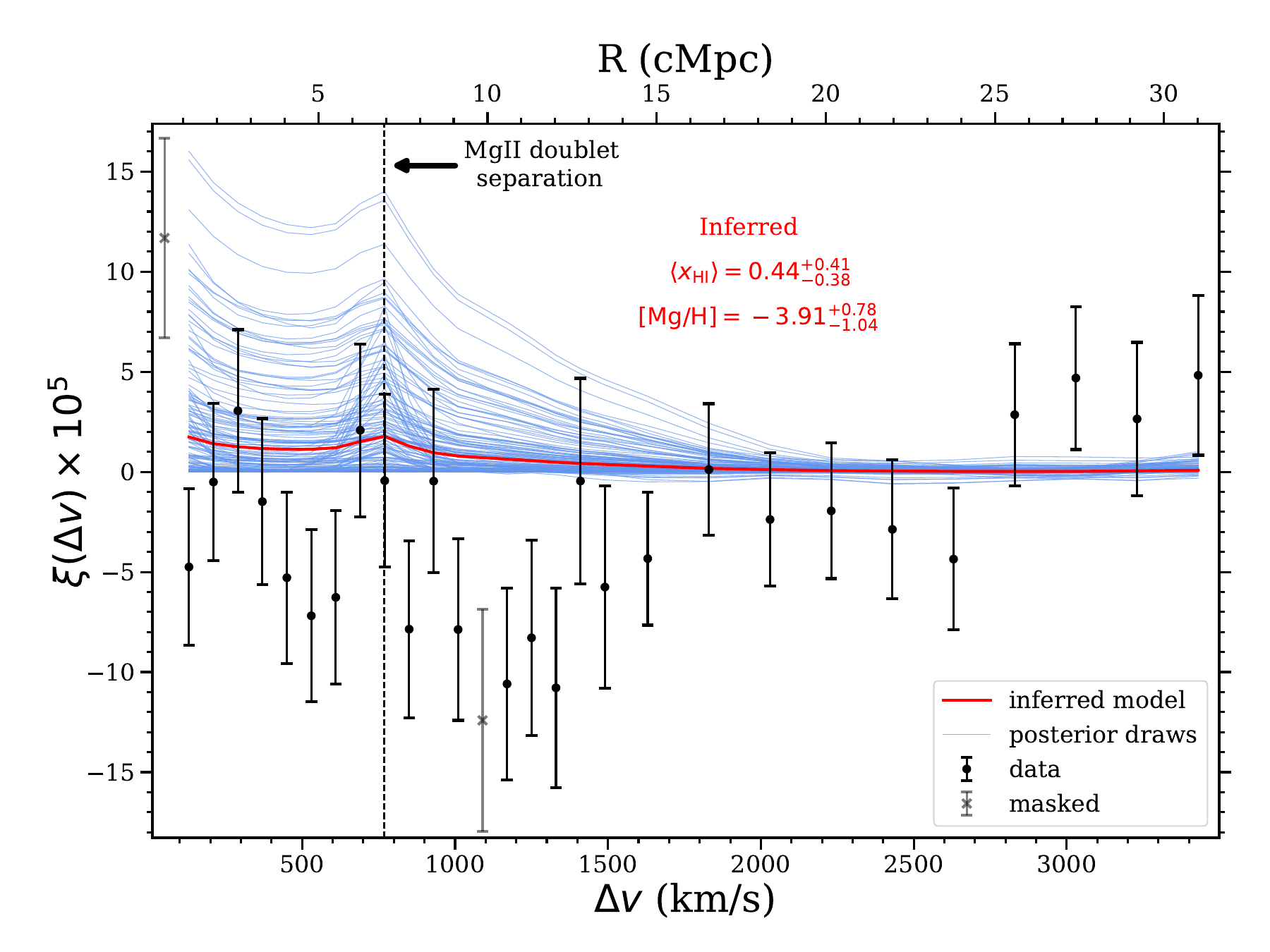}
\includegraphics[height=0.29\textheight, width=0.37\textwidth]{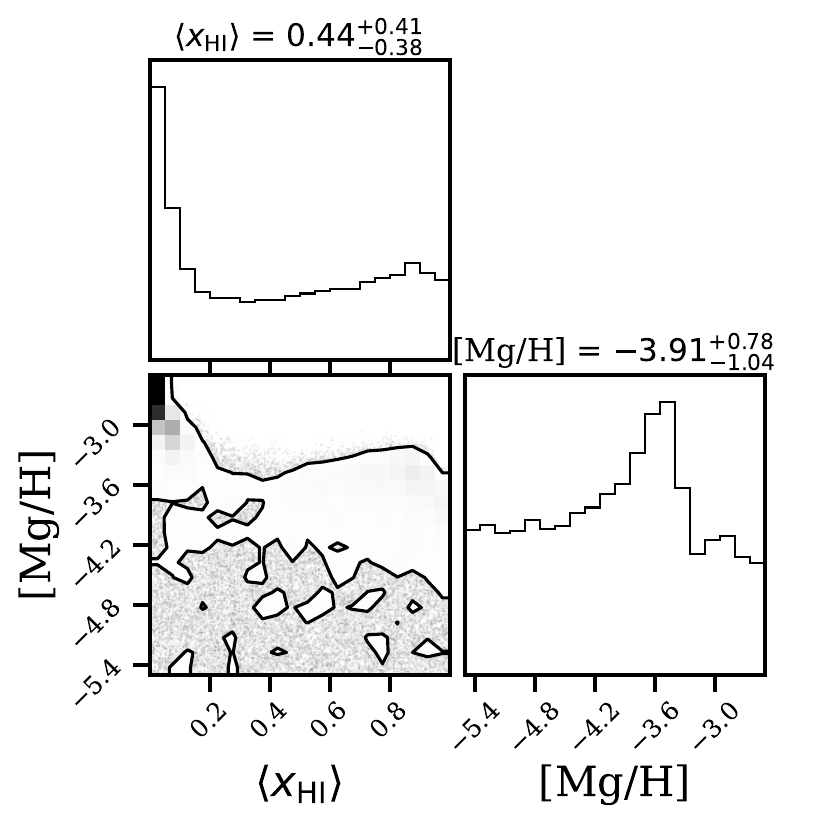}
\caption{Same as Figure \ref{allz-mcmc} but for the high-$z$ bin ($z_{\rm{\ion{Mg}{II},med}} = 6.715$). The 95\% C.L. upper limit on [Mg/H] is $-3.45$. }
\label{highz-mcmc}\
\end{figure*}

\begin{figure*}
\includegraphics[height=0.35\textheight, width=0.62\textwidth]{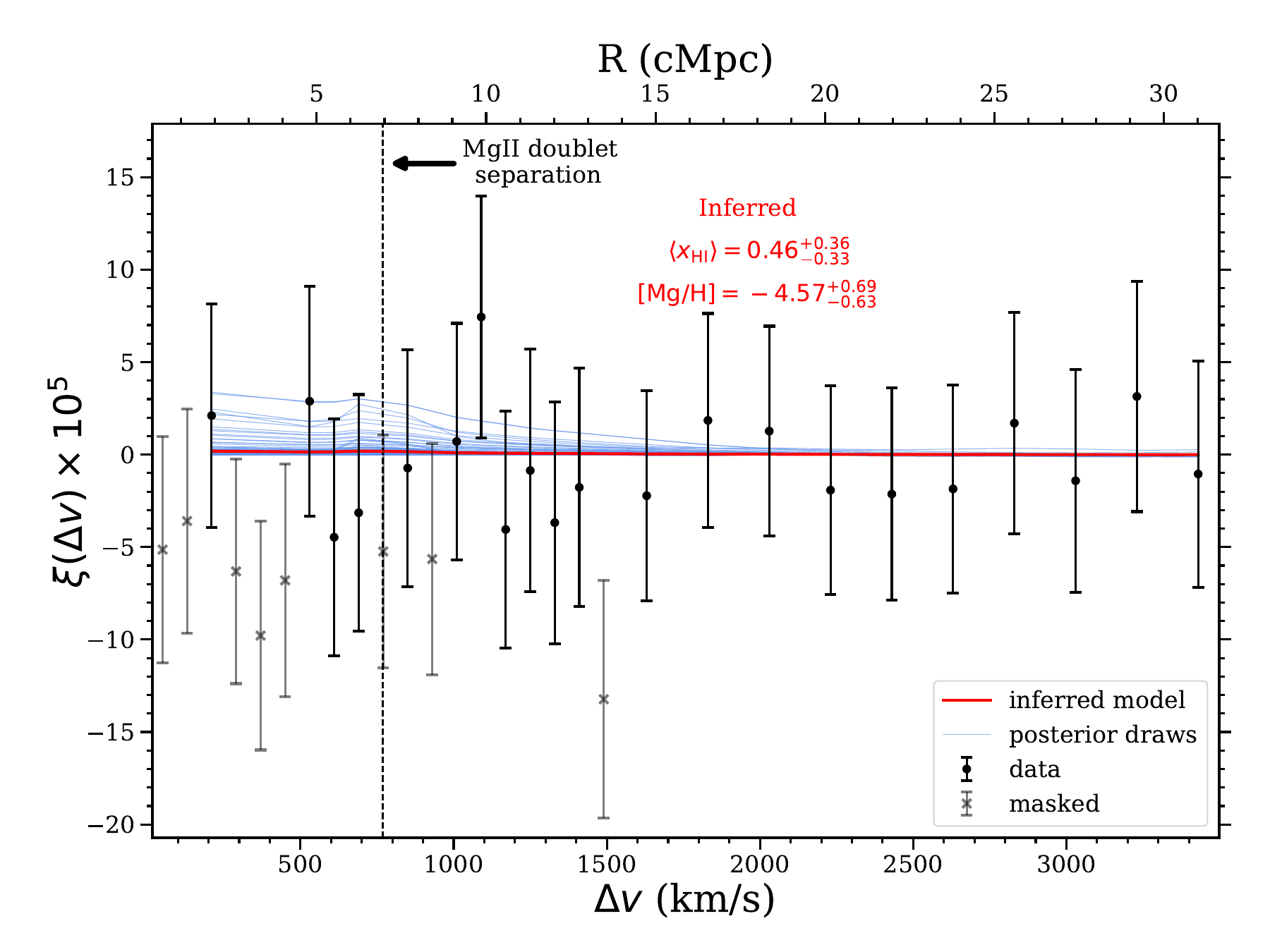}
\includegraphics[height=0.29\textheight, width=0.37\textwidth]{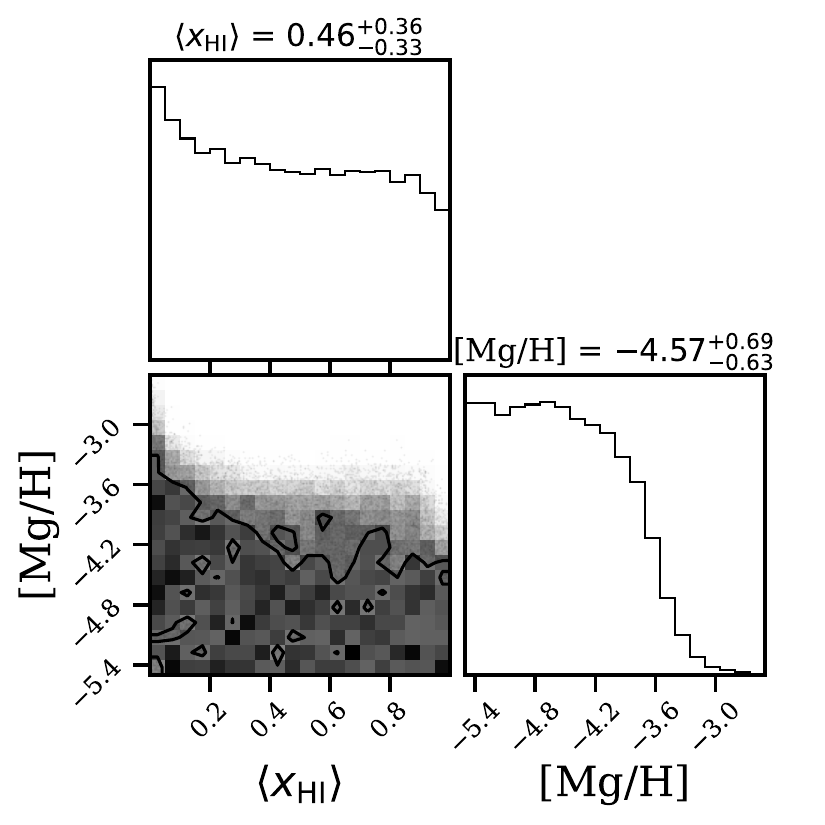}
\caption{Same as Figure \ref{allz-mcmc} but for the low-$z$ bin ($z_{\rm{\ion{Mg}{II},med}} = 6.235$). The 95\% C.L. upper limit on [Mg/H] is $-3.75$.}
\label{lowz-mcmc}\
\end{figure*}

\section{Conclusions}
The auto-correlation of the \ion{Mg}{II} forest, when treated as a continuous random  field, has been shown to probe the neutral fraction and metallicity of the IGM (\citetalias{Hennawi2021}). We measure this statistic for the first time using a sample of ten $z \geq 6.80$ quasars observed with the Keck/MOSFIRE, Keck/NIRES, and VLT/X-SHOOTER spectrographs, where the median redshift of our measurement is $z = 6.469$. We also measure the correlation function over three redshift bins: ``low-$z$'' ($z_{\rm \ion{Mg}{II}} < 6.469$, $z_{\rm{\ion{Mg}{II},med}} = 6.235$), ``high-$z$'' ($z_{\rm \ion{Mg}{II}} \geq 6.469$, $z_{\rm{\ion{Mg}{II},med}} = 6.715$), and ``all-$z$'' ($z_{\rm{\ion{Mg}{II},med}} = 6.469$). Our masking procedures to remove CGM absorbers successfully recover known strong absorbers as well as weaker absorbers. 

\begin{figure}
\includegraphics[trim=20 20 20 10,clip,width=\columnwidth, height=0.63\columnwidth]{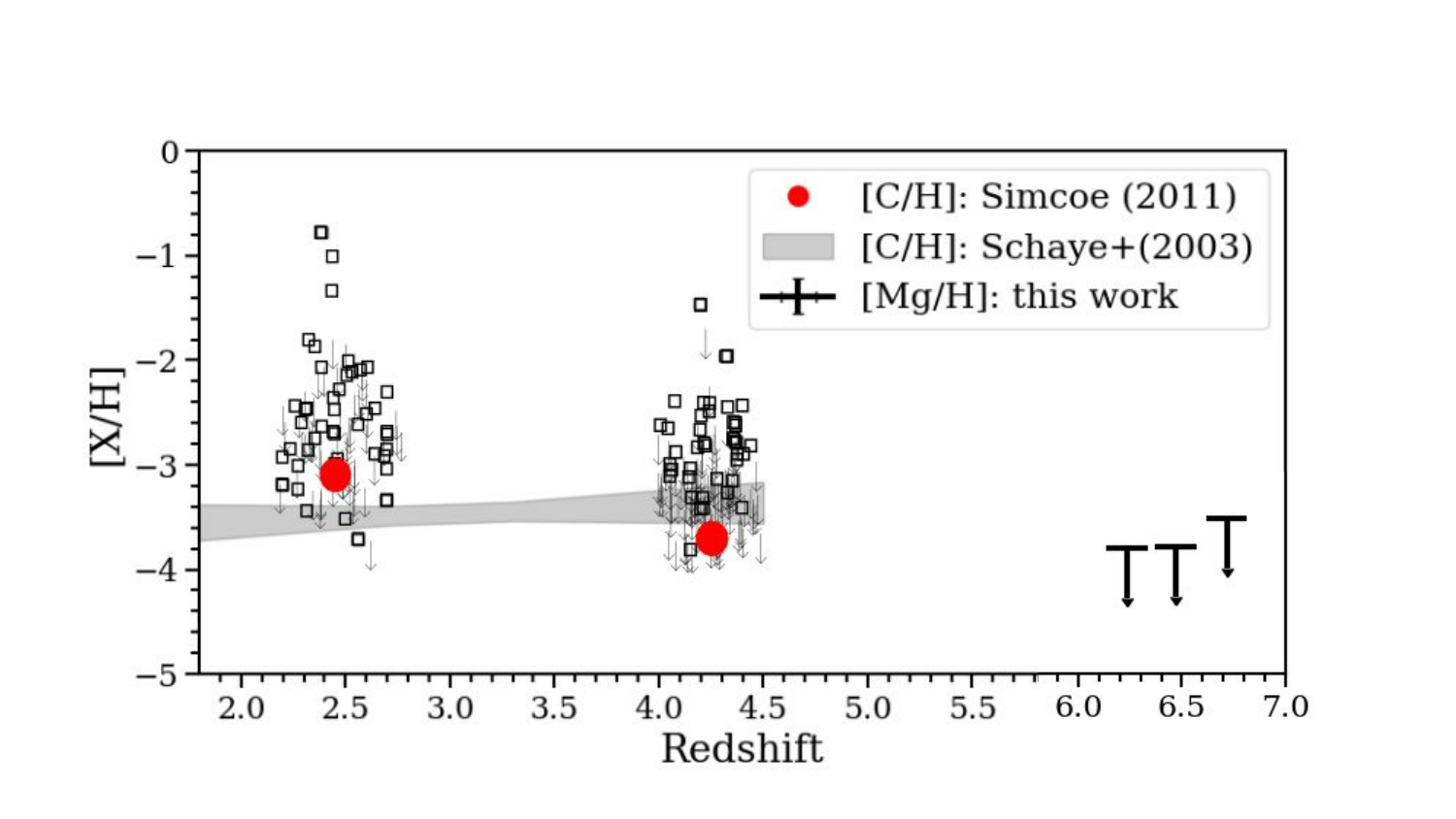}
\vspace{-4ex}
\caption{Measurements of the IGM metallicity as a function of redshift. Both {\protect\cite{Simcoe2011a}} and {\protect\cite{Schaye2003}} use \ion{C}{IV} as their metal ion of choice. Our upper limits from this work are shown in black arrows, where from left to right, ``low-$z$'' (at $z_{\rm{\ion{Mg}{II},med}}$ = 6.235), ``all-$z$'' (at $z_{\rm{\ion{Mg}{II},med}}$ = 6.469), and ``high-$z$'' (at $z_{\rm{\ion{Mg}{II},med}}$ = 6.715). This figure is adapted from Figure 16 of {\protect\cite{Simcoe2011a}} using the best-fit measurement from {\protect\cite{Schaye2003}}.}
\label{evol}
\end{figure}

The correlation function of the \ion{Mg}{II} forest without the CGM absorbers masked validates our measurement technique. 
We detect strong peaks  in the correlation functions at the characteristic velocity separation of the \ion{Mg}{II} doublet (768 km/s)  in all redshift bins as well as a rise towards small velocity lags owing to the velocity structure of CGM absorbers. After masking out the identified CGM absorbers, our correlation function measurements yield a null result that is consistent with noise in all redshift bins. We performed Bayesian inference where forward-modeled spectra were used to syntehsize a model-dependent covariance matrix of the data. Our MCMC analysis yields an upper limit of [Mg/H] $< -3.73$ at 95\% confidence at a median redshift $z_{\rm \ion{Mg}{II}} = 6.47$, while we are unable to place any strong constraints on the hydrogen neutral fraction. We place our upper limits in context with other literature measurements of the IGM metallicity in Figure \ref{evol}. \cite{Simcoe2011a} (red points in figure) obtained an estimate of [C/H] via Voigt profile fitting of a sample of discrete \ion{H}{I} absorbers with log$(\rho/\bar{\rho}) = 0.47$, finding a median abundance [C/H] = $-3.55$ at $z \sim 4.3$, which is $2-3$ times lower than similar measurements at $z \sim 2.4$ using both \ion{C}{IV} and \ion{O}{VI}.  On the other hand, \cite{Schaye2003} measured the distribution of \ion{C}{IV} pixel optical depth as a function of the corresponding \ion{H}{I} optical depth and converted this relation into an estimate of [C/H] as a function of density (the so-called pixel optical method). They measured a median [C/H] $= -3.47$ at $z = 3$ in the low density IGM gas with log$(\rho/\bar{\rho}) = 0.5$ and found little evolution in [C/H] from $z=4$ to $z=2$, as indicated by their best-fit model in the gray shaded region. To establish the existence or lack thereof of an evolution in the IGM metallicity, it is crucial to perform more measurements at $z \sim 3$ and fill the redshift gap at $z > 4.5$ and our results provide an important step in this direction. 

Despite a null detection in this first study, a statistical method like the correlation function remains one of the best ways to measure the IGM metallicity at high redshift, which has never been attempted before due to observational difficulties and challenges posed by the traditional method of discrete line identification. Looking ahead to the future with spectra from JWST (James Webb Space Telescope; \citealp{JWST2006}), we expect to place stronger constraints on primordial cosmic enrichment and reionization. For instance, using a sample of eight $z \geq 7$ quasar spectra with a SNR/pixel of $\sim$76 from JWST/NIRSpec\footnote{\url{https://www.stsci.edu/jwst/science-execution/program-information?id=3526}}, we expect to jointly constrain [Mg/H] within 1$\sigma$ precision of 0.04 dex and $x_{\rm{\ion{H}{I}}}$ within 1$\sigma$ precision of 10\%, as well as place an upper limit of [Mg/H] $< -3.94$ on the IGM enrichment at 95\% credibility. Finally, by combining very high-$z$ measurements from JWST that is probing the neutral IGM with ground-based measurements of, e.g. the \ion{C}{IV} forest, at $z \sim 3-4$ that is probing the \textit{ionized} IGM, we may be able to detect the transition from forests of low-ionization metal absorbers to forests of high-ionization lines, which is the smoking gun of reionization as seen through metal absorption lines.


\label{conclusion}

\section*{Acknowledgment}
We acknowledge helpful conversations with the ENIGMA group at UC Santa Barbara and Leiden University. This project has received funding from the European Research Council (ERC) under the European Union’s Horizon 2020 research and innovation program (grant agreement No 885301). JFH acknowledges support from the National Science Foundation under Grant No. 1816006. JO acknowledges support from grants PID2022-138855NB-C32 and CNS2022-135878 from the Spanish Ministerio de Ciencia y Tecnologia. This research made use of \texttt{Astropy}\footnote{\url{http://www.astropy.org}}, a community-developed core Python package for Astronomy \citep{Astropy2013,Astropy2018} and \texttt{SciPy}\footnote{\url{https://scipy.org}} \citep{Scipy2020}.

\section*{Data availability}
The data underlying this article will be shared on reasonable request to the corresponding author.

\section*{Appendix A}
During our investigations, we discovered that our likelihood yields MCMC samples and best-fit models that appear visually inconsistent with the actual measurements and error bars. The source of this discrepancy is correlated velocity bins with negative values in the data that are not supposed to occur given the correlation structure of the covariance matrix.
To rectify this, 
we mask the outlying velocity bins before performing the MCMC. To determine which bins should be masked, we compare values of the correlation function measured from the data with values measured from the forward models in which the IGM model has no detectable signal. This is because which velocity bins are outliers depend on the structure of the covariance matrix, which is model-dependent. Therefore, we choose a model with no signal to remove sensitivity towards the signal.

\begin{figure*}
\centering
\includegraphics[width=\columnwidth]{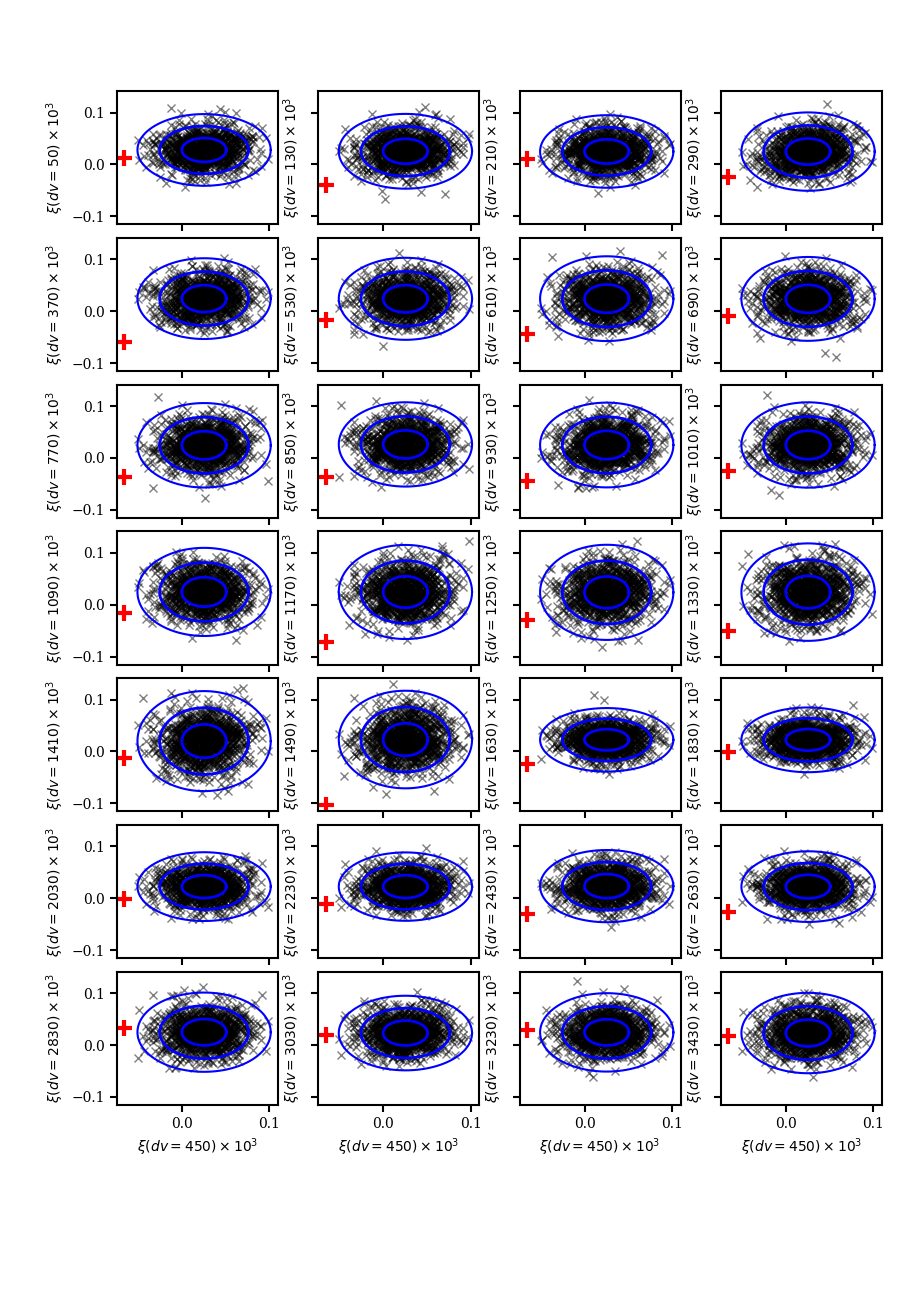}\centering
\includegraphics[width=\columnwidth]{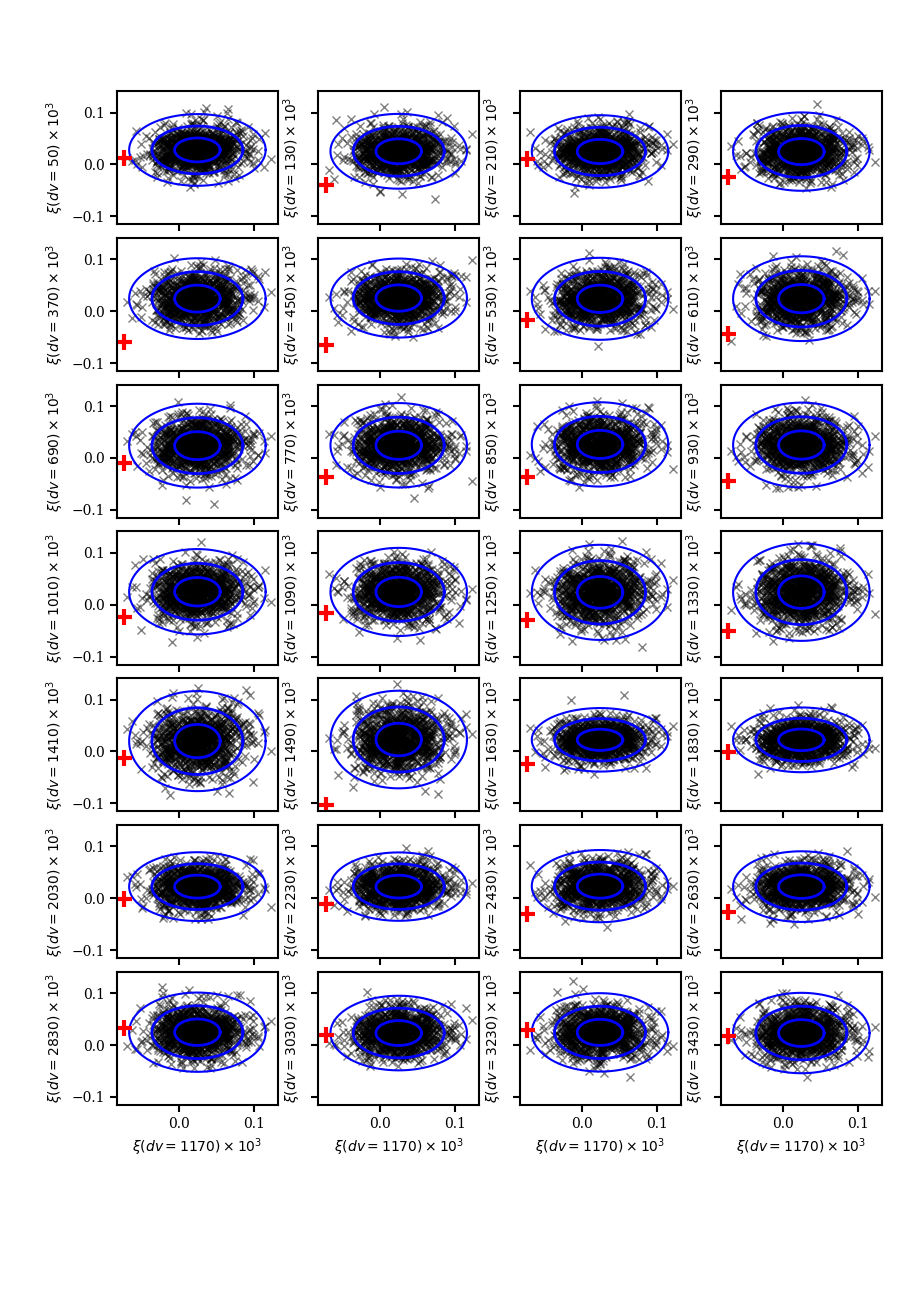}
\caption{Values of the all-$z$ redshift bin correlation function at $dv=450$ km/s (left) and $dv=1170$ km/s (right) plotted against the values of all other bins. The black points are computed from forward models with $x_{\rm{\ion{H}{I}}}$ = 0.50 and [Mg/H] = $-4.50$ and the ellipses denote their 1, 2, and 3$\sigma$ contours. Red points are measurements from the real data that fall outside of the 3$\sigma$ ellipse, where these two bins are examples of catastrophic failure as the real data fall outside the 3$\sigma$ ellipse of all the other bins.}
\label{corrbin-bad}
\end{figure*}

Figures \ref{corrbin-bad} and \ref{corrbin-good} show the correlation functions in one velocity lag bin plotted against the values of all other velocity bins for the all-$z$ bin. The black points are measured from forward models with $x_{\rm{\ion{H}{I}}}$ = 0.50 and [Mg/H] = $-4.50$, where the ellipses denote their 1, 2, and 3$\sigma$ contours. Measurements from the real data that fall outside of the 3$\sigma$ ellipse are marked as red while measurements that fall within the 3$\sigma$ ellipse are marked as green. 
We mask the velocity bin where 20 or more of the real measurements fall outside the 3$\sigma$ ellipse; Figure \ref{corrbin-bad} shows two examples of bad bins that are masked, while Figure \ref{corrbin-good} shows two examples of good bins that we retain. In conclusion, we mask $dv$ = 370, 450, 1170, and 1490 km/s in the all-$z$ bin, $dv$ = 50 and 1090 km/s in the high-$z$ bin, and $dv$ = 50, 130, 290, 370, 450, 770, 930, and 1490 km/s in the low-$z$ bin (see Figures \ref{allz-mcmc} - \ref{lowz-mcmc}).

We generate the forward-modeled datasets and compute their covariance matrices for all bins and only apply the bin masking before the MCMC sampling, in which we mask the bad bins in both the correlation function and the covariance matrix. The best-fit models and upper limits are therefore obtained from the masked data and masked covariance matrices. 

\begin{figure*}
\centering
\includegraphics[width=\columnwidth]{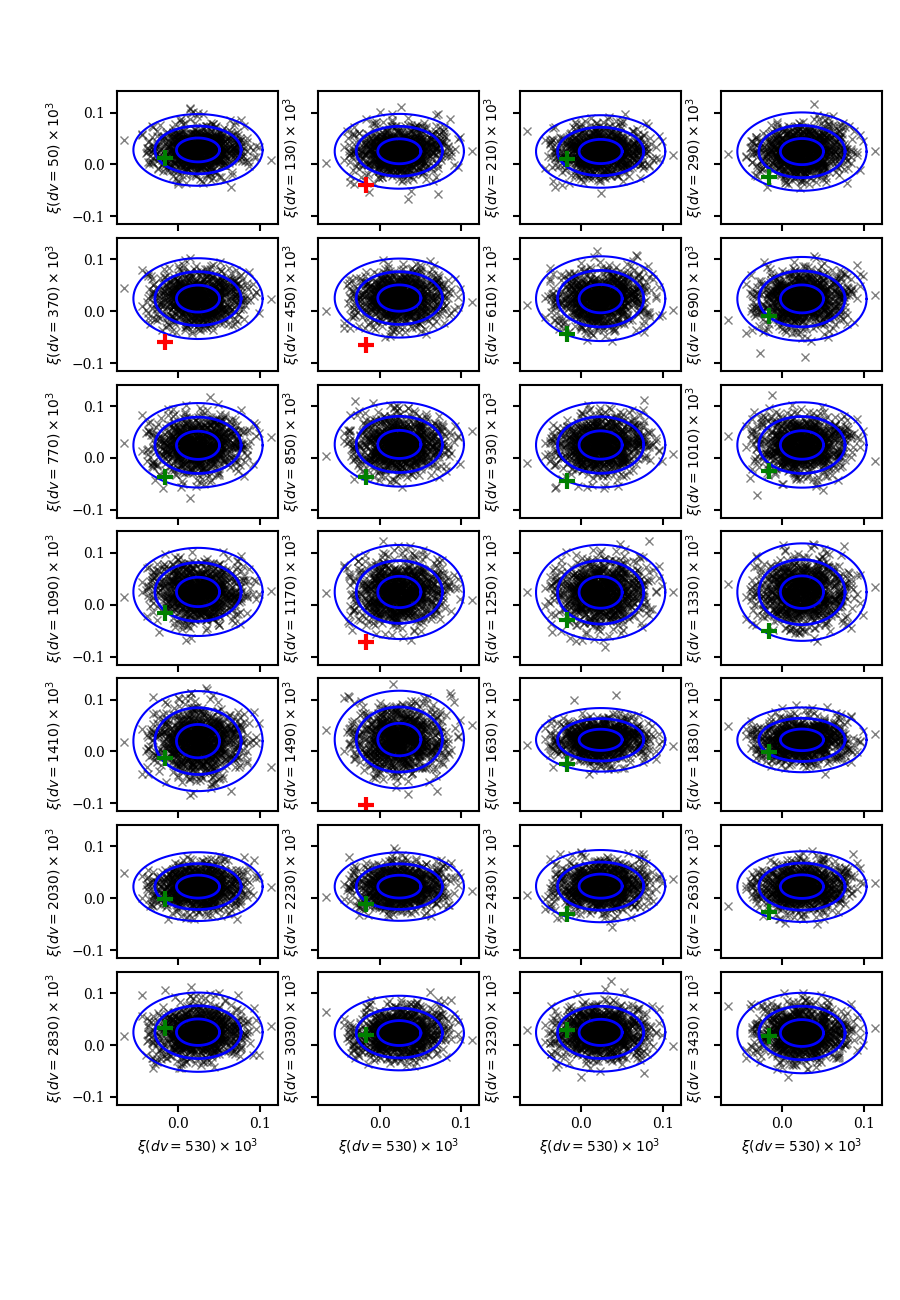}\centering
\includegraphics[width=\columnwidth]{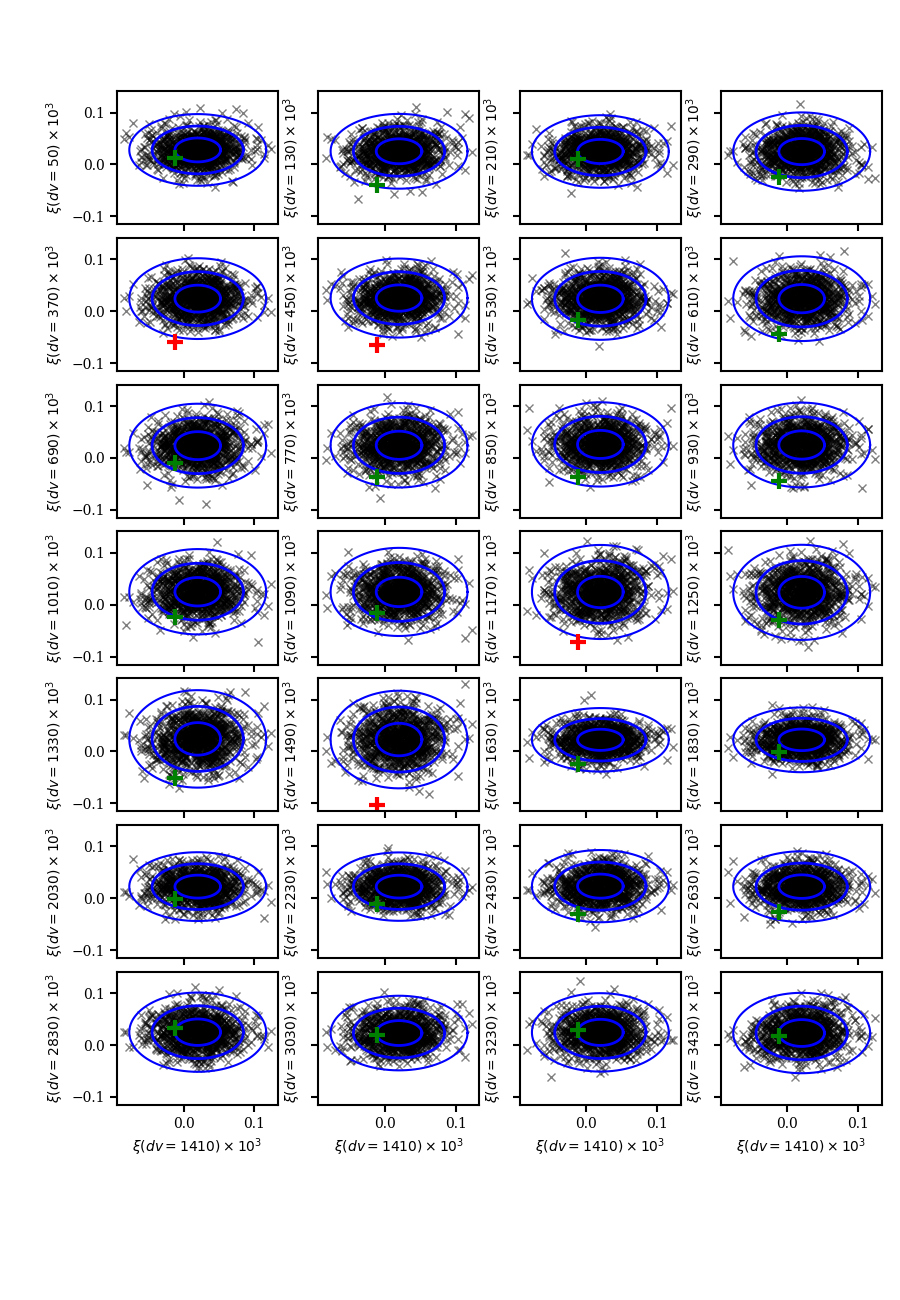}
\caption{Same as Figure \ref{corrbin-bad}, but for $dv=530$ km/s (left) and $dv=1410$ km/s (right). Red (green) points are measurements from the real data that fall outside of (within) the 3$\sigma$ ellipse. In these two bins, most the real data fall within the 3$\sigma$ ellipse of the other bins.}
\label{corrbin-good}
\end{figure*}

\section*{Appendix B}
Here we show the masked absorbers for the rest of the quasars in our dataset that are not shown in \S \ref{masking}, which are Figure \ref{masked-allother-qso} and \ref{masked-allother-qso2}.

\begin{figure*}
\includegraphics[width=\textwidth]{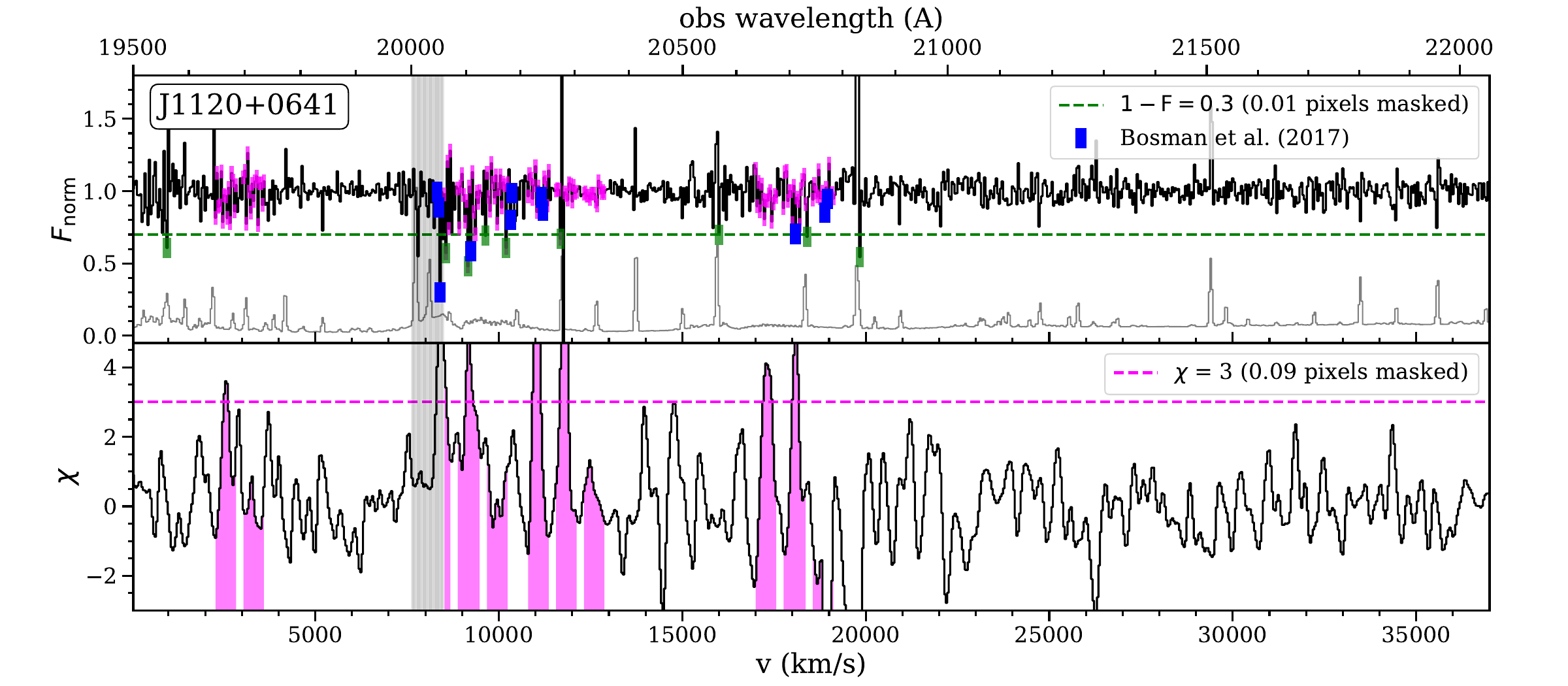}
\includegraphics[width=\textwidth]{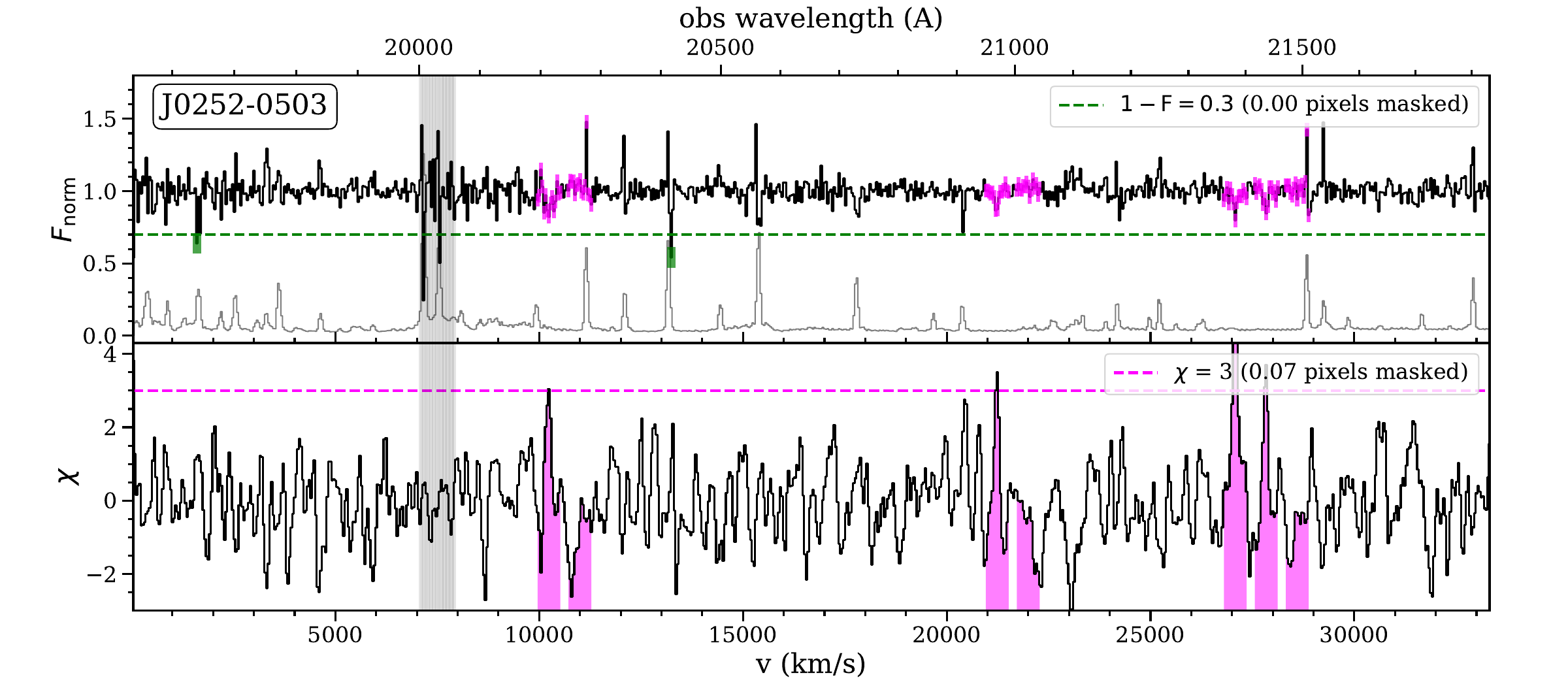}
\includegraphics[width=\textwidth]{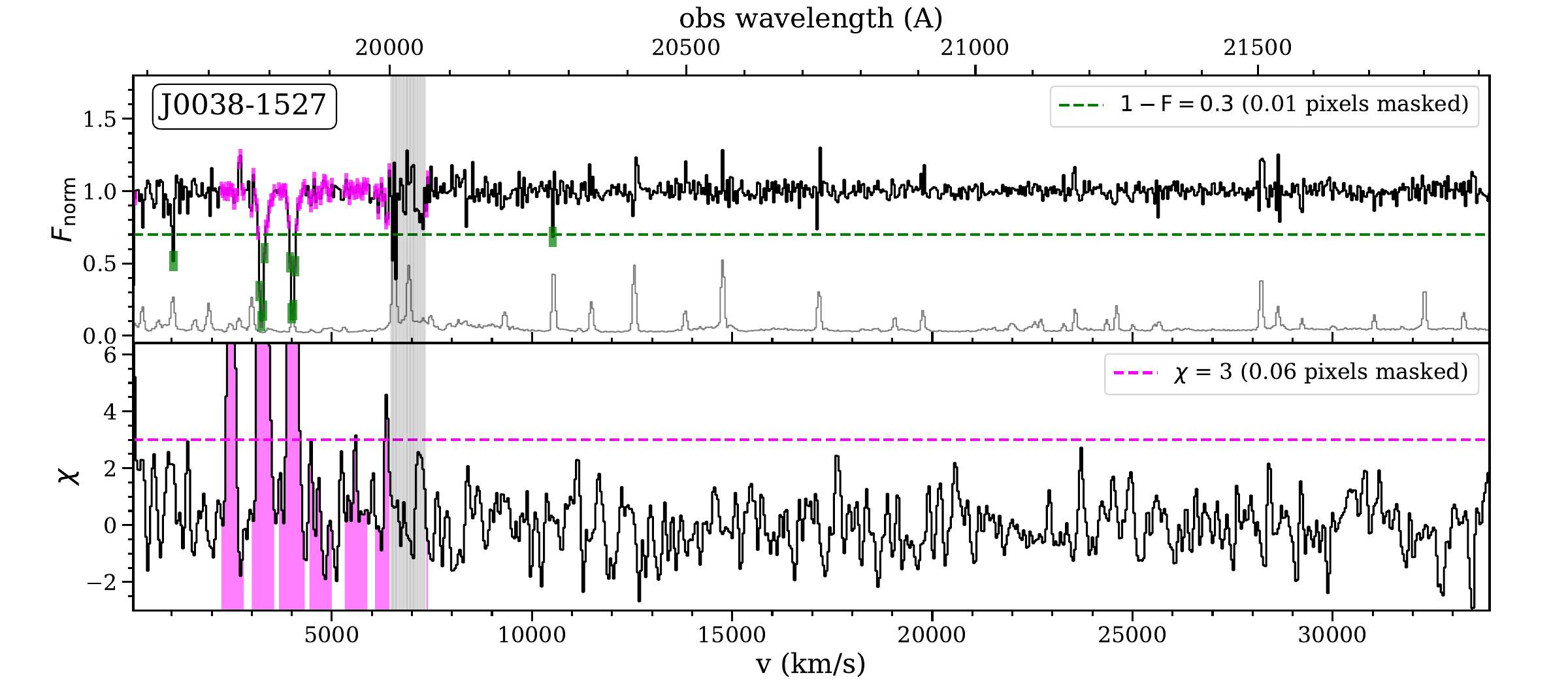}
\vspace{-2ex}
\caption{Continued from Figure \ref{masked-all-qso}, for J1120$+$0641 (top), J0252$-$0503 (middle), and J0038$-$1527 (bottom). For J1120$+$0641, we also manually mask the locations of weak absorbers found by {\protect\cite{Bosman2017}}, as indicated by the blue ticks.}
\label{masked-allother-qso}
\end{figure*}

\begin{figure*}
\includegraphics[width=\textwidth]{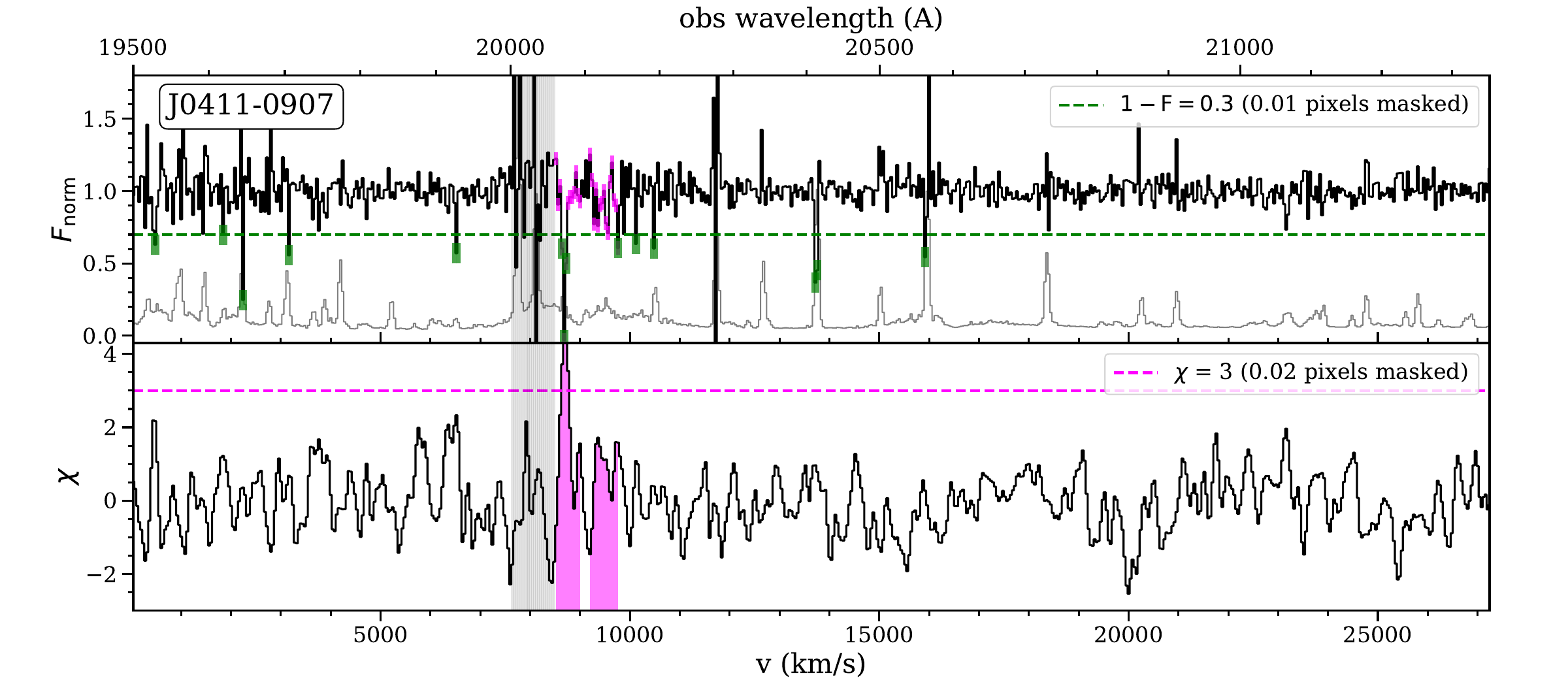}
\includegraphics[width=\textwidth]{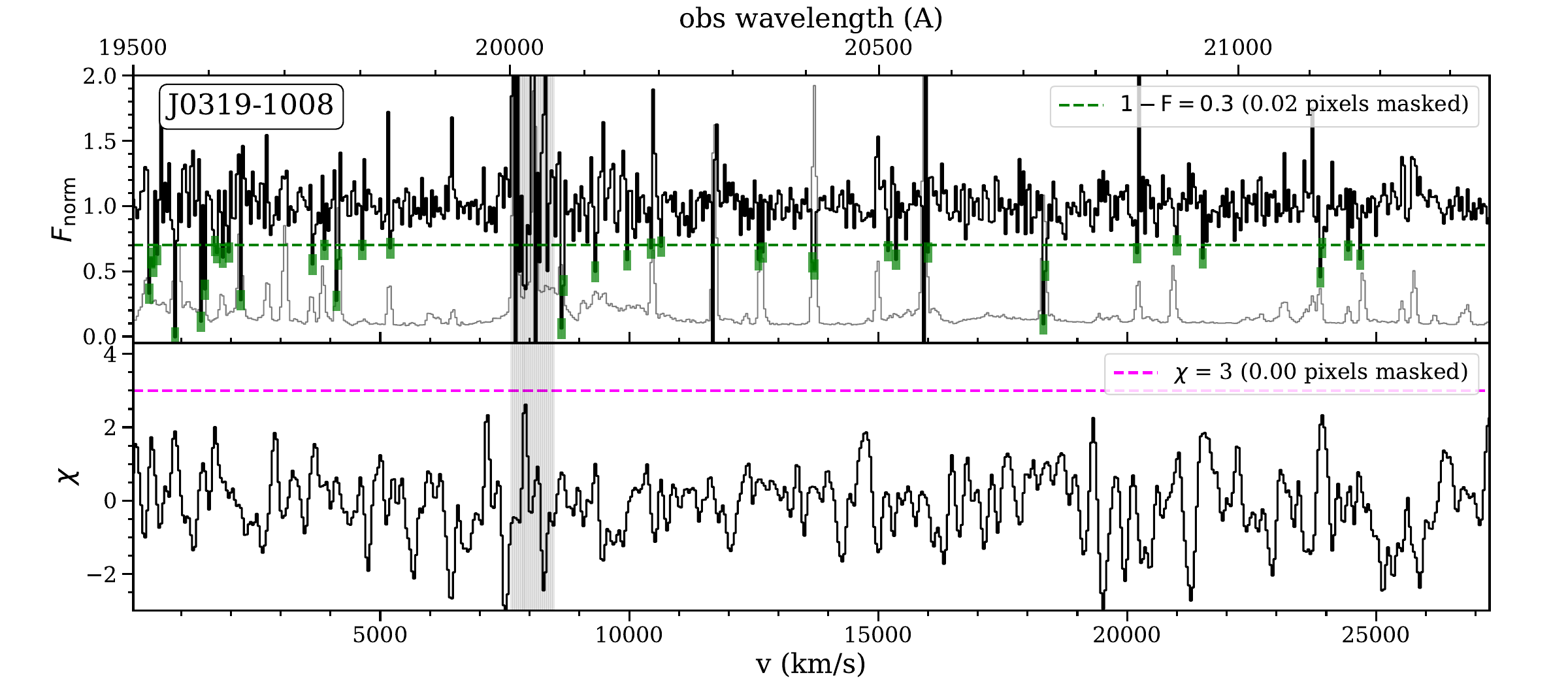}
\caption{Continued from Figure \ref{masked-all-qso}, for J0411$-$0907 (top) and J0319$-$1008 (bottom).}
\label{masked-allother-qso2}
\end{figure*}

\section*{Appendix C}
Here we show the forward-modeled spectra for the rest of the quasars in our dataset that are not shown in \S \ref{results}, assuming an IGM model with (${x_{\rm{\ion{H}{I}}}, \rm{[Mg/H]}}) = (0.50, -4.50)$. These are shown in Figure \ref{forwardmodels2} and \ref{forwardmodels3}.

\begin{figure*}
\includegraphics[width=\textwidth]{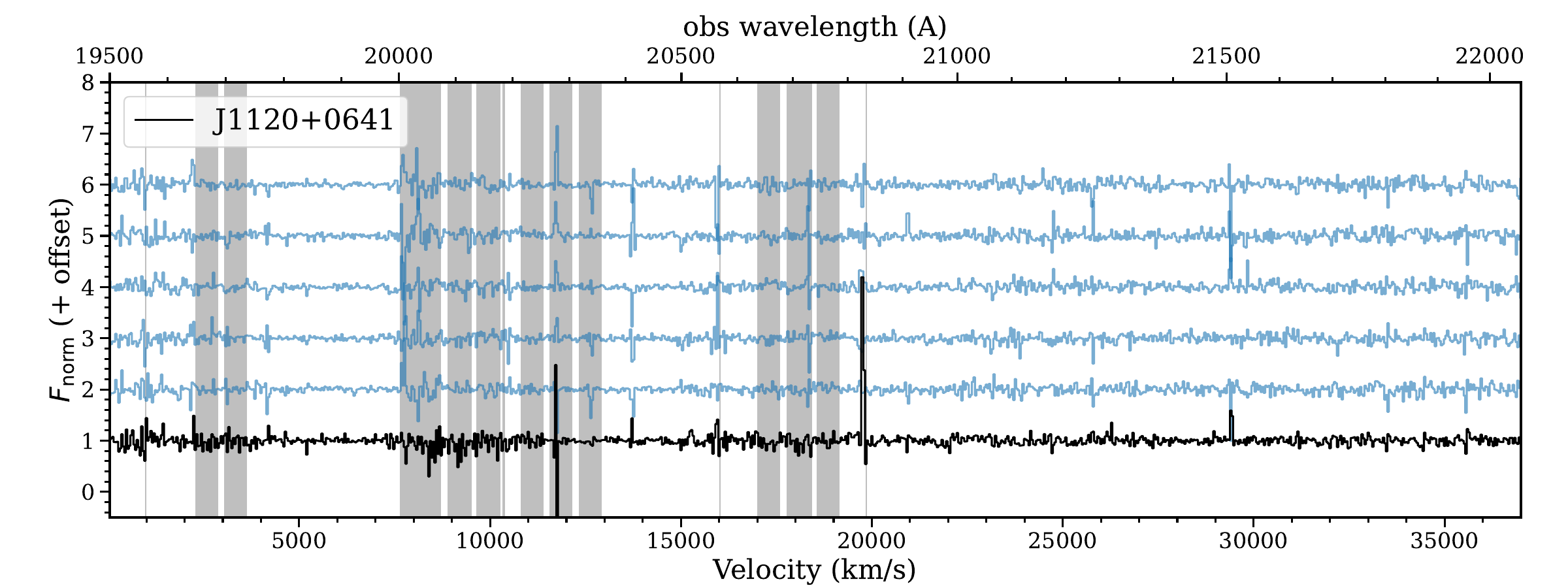}
\includegraphics[width=\textwidth]{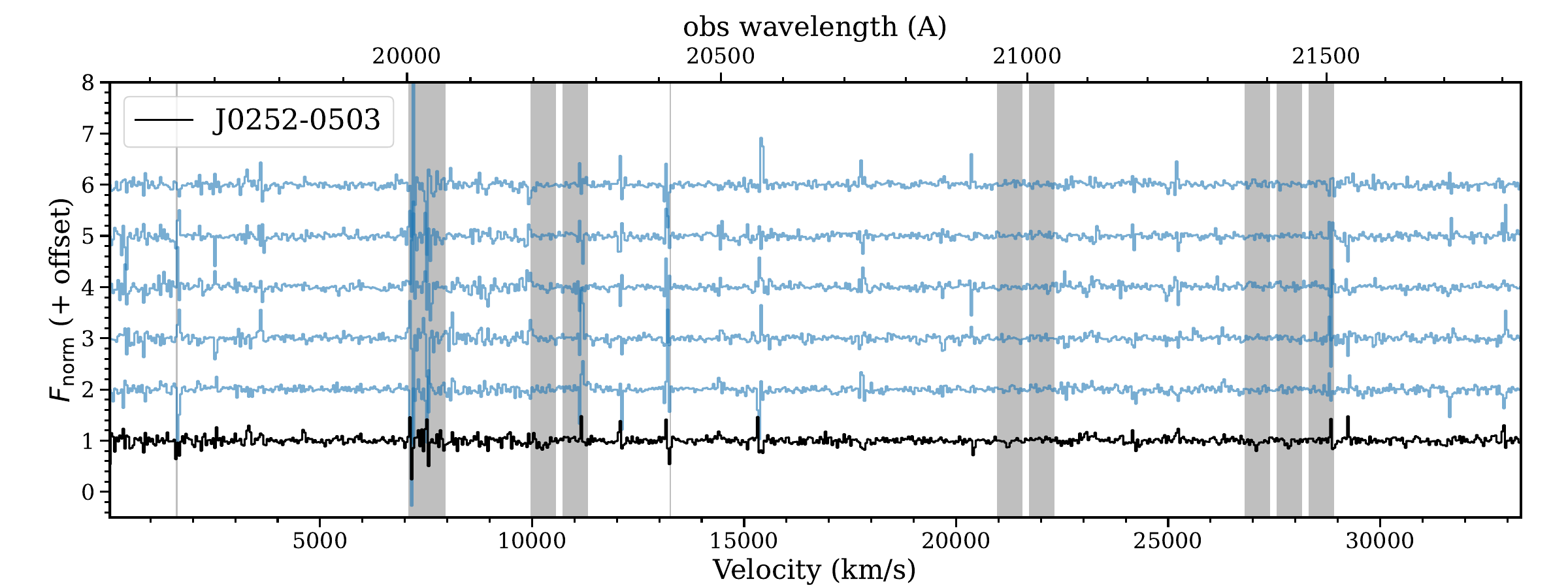}
\includegraphics[width=\textwidth]{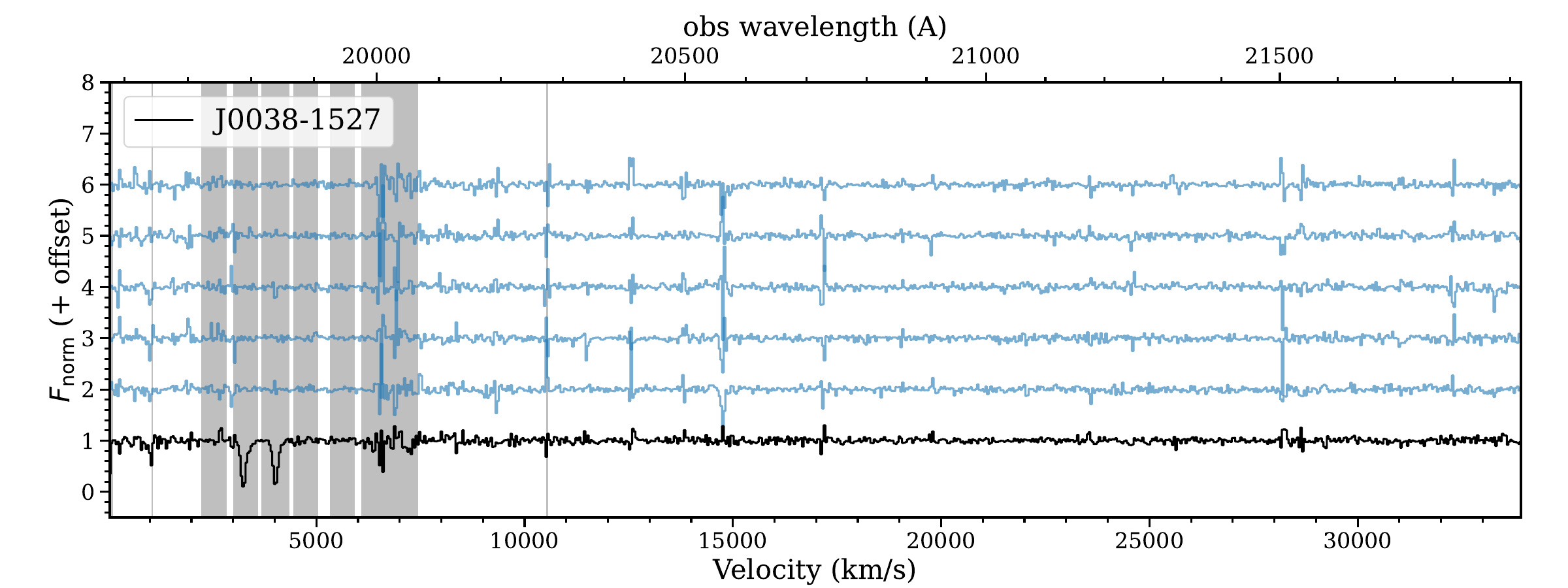}
\caption{Continued from Figure \ref{forwardmodels}, for J1120$+$0641 (top), J0252$-$0503 (middle), and J0038$-$1527 (bottom).}
\label{forwardmodels2}
\end{figure*}

\begin{figure*}
\includegraphics[width=\textwidth]{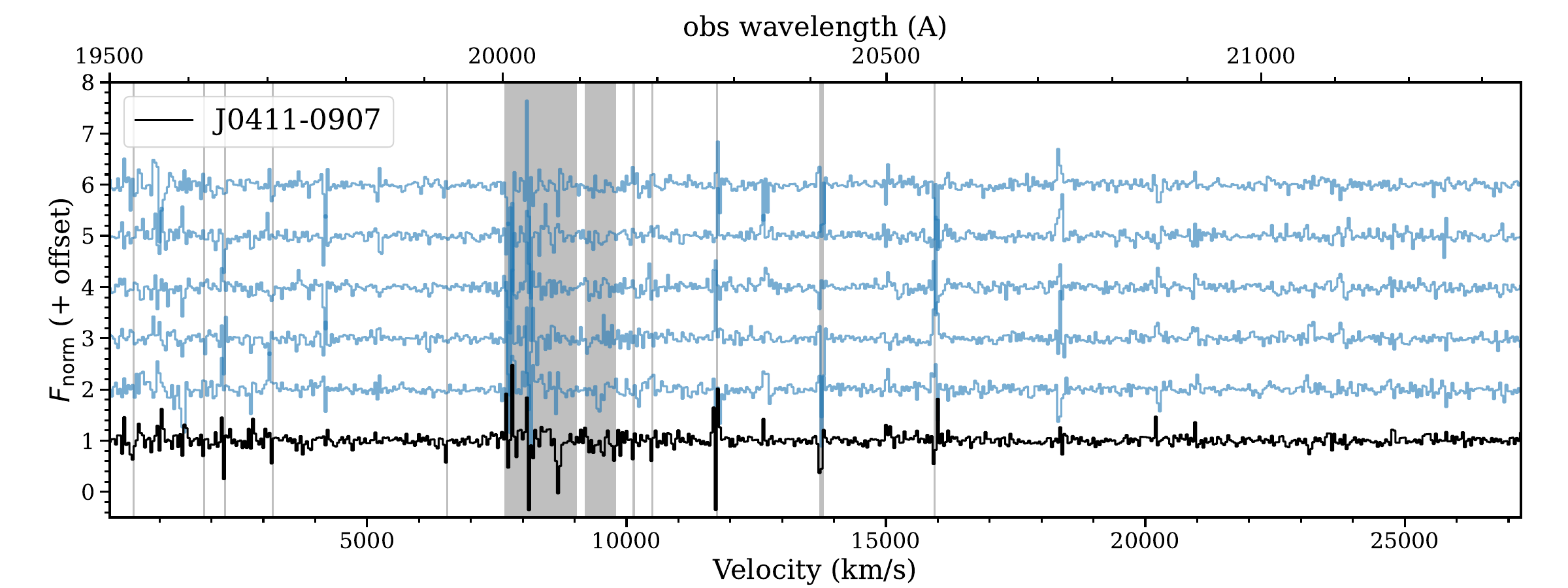}
\includegraphics[width=\textwidth]{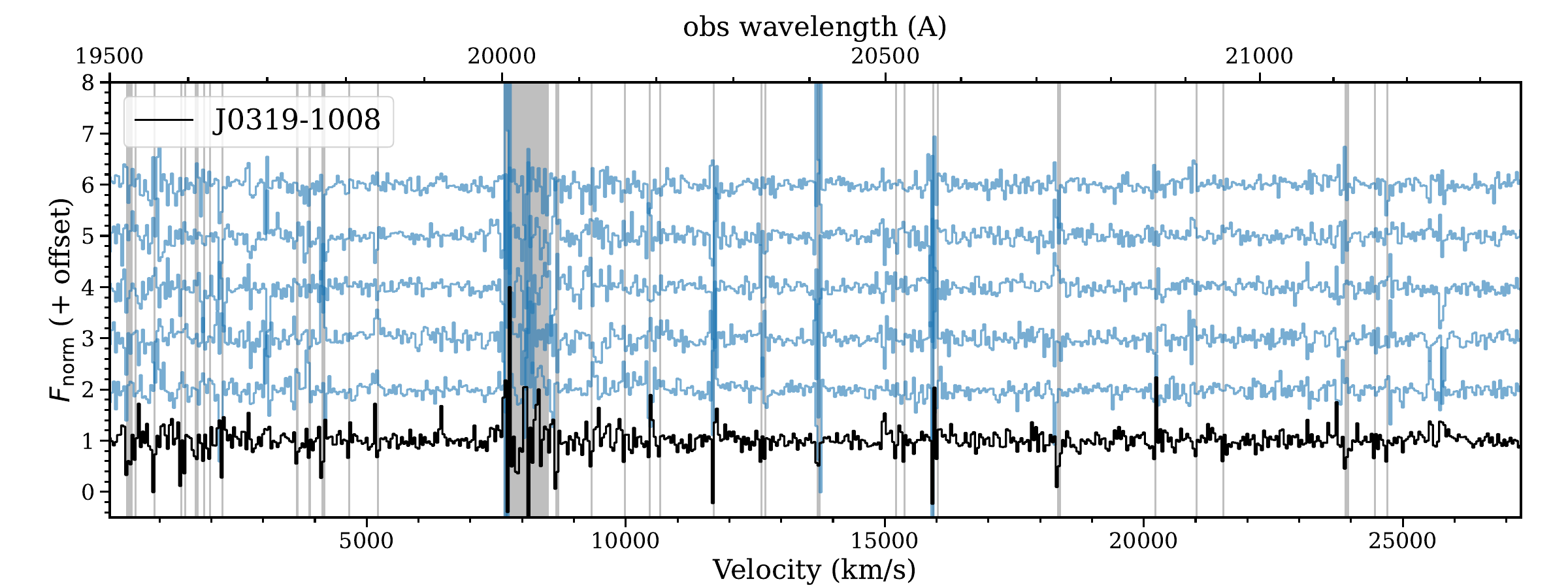}
\caption{Continued from Figure \ref{forwardmodels2}, for J0411$-$0907 (top) and J0319$-$1008 (bottom).}
\label{forwardmodels3}
\end{figure*}

\end{document}